\def\aj{{AJ}}

\def\apj{{ApJ}}

\def\col#1{\empty} 
\def\bw#1{#1}       % Black-and-white figures

\def\col#1{#1}      % Color figures
\def\bw#1{\empty}
                                                            
\newdimen\sa  \def\sd{\sa=.1em  \ifmmode $\rlap{.}$''$\kern -\sa$
                                \else \rlap{.}$''$\kern -\sa\fi}
\def\ts{\thinspace}
\def\etal{{et~al.~}} 
\def\gapprox{$_>\atop{^\sim}$}
\def\lapprox{$_<\atop{^\sim}$}

\documentclass[cup5b]{caps}

\usepackage{graphicx}
\usepackage{ociwsymp1}
\HeadText{J. Kormendy}

\begin{document}

\pagenumbering{arabic}

\author[]{JOHN KORMENDY\\Department of Astronomy, University of Texas at Austin}

% Luis, the title wraps most neatly if I write it like this:

\chapter{The Stellar-Dynamical Search ~~~~~~~~~~~~~
         for Supermassive Black Holes ~~~~~~~~~~~~~~
         in Galactic Nuclei}

\vskip -10pt

\begin{abstract}

      The robustness of stellar-dynamical black hole (BH) mass measurements is
\hbox{illustrated} using six galaxies that have results from independent
research groups.  Derived BH masses have remained constant to a factor of 
$\sim 2$ as spatial resolution has improved by a factor of \hbox{2 -- 330,}
as velocity distributions have been measured in increasing detail, and as the
analysis has improved from spherical, isotropic models to axisymmetric, 
three-integral models.  This gives us confidence that the masses are reliable
and that the galaxies do not indulge in a wide variety of perverse orbital
structures.  Another successful test is the agreement between a preliminary
stellar-dynamical BH mass for NGC 4258 and the accurate mass provided by the
maser disk.  Constraints on BH alternatives are also improving.  In
M{\thinspace}31, {\it Hubble Space Telescope\/} ({\it HST\/}) spectroscopy 
shows that the central massive dark object (MDO) is in a tiny cluster of blue
stars embedded in the P2 nucleus of the galaxy.  The MDO must have a radius 
$r$ \lapprox \ts0\sd06.  M{\thinspace}31 becomes
the third galaxy in which dark clusters of brown dwarf stars or stellar remnants
can be excluded.  In our Galaxy, spectacular proper motion observations of
almost-complete stellar orbits show that the central dark object has radius $r$
\lapprox \ts0.0006 pc. Among BH alternatives, this excludes even neutrino balls.
Therefore, measurements of central dark masses and the conclusion that these are
BHs have both stood the test of time.  Confidence in the BH paradigm for active
galactic nuclei (AGNs) is correspondingly high.

      Compared to the radius of the BH sphere of influence, BHs are
being discovered at similar spatial resolution with {\it HST\/}
as in ground-based work. The reason is that {\it HST\/} is used to
observe more distant galaxies.  Typical BHs are detectable in the Virgo
cluster, and the most massive ones are detectable 3 -- 6 times farther
away.  Large, unbiased samples are accessible.  As a result, {\it
HST\/} has revolutionized the study of BH demographics.  

\end{abstract}

\section{Introduction}

      The supermassive black hole paradigm for AGNs was launched by 
Zel'dovich (1964), Salpeter (1964), and Lynden-Bell (1969, 1978), who argued
that the high energy production efficiencies required to make
quasars are provided by gravity power.  Eddington-limited accretion suggested 
that BH engines have masses of $10^6$ to $10^9$ $M_\odot$.  Confidence grew 
rapidly with the amazing progress in AGN observations and with the paradigm's
success in weaving these results into a coherent theoretical picture.  Unlike
the normal course of scientific research, acceptance of the AGN paradigm came
long before there was any dynamical evidence that BHs exist. 

\vfill\eject

\centerline{\null} \vfill

\includegraphics{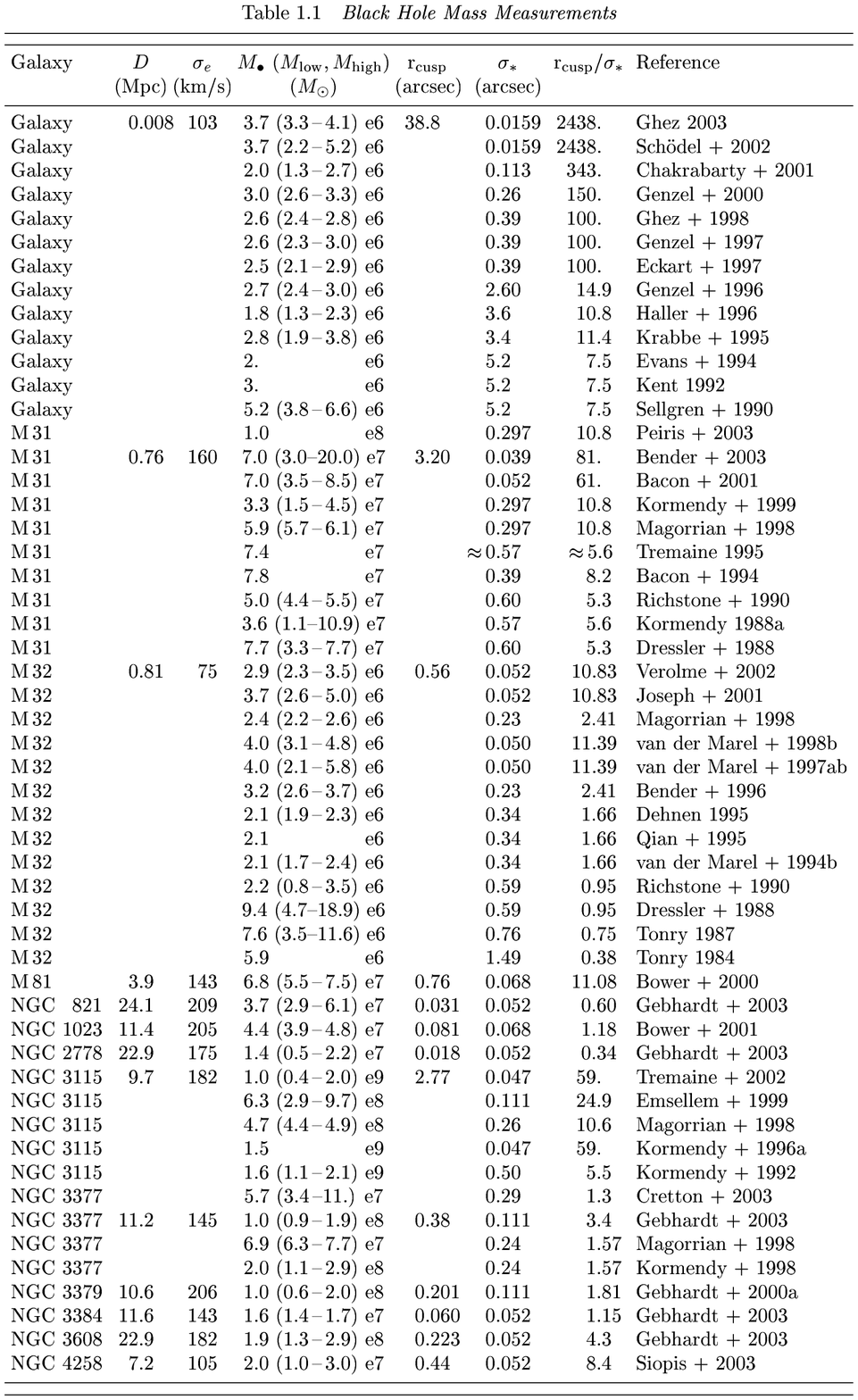}

\eject

\centerline{\null} \vfill

\includegraphics{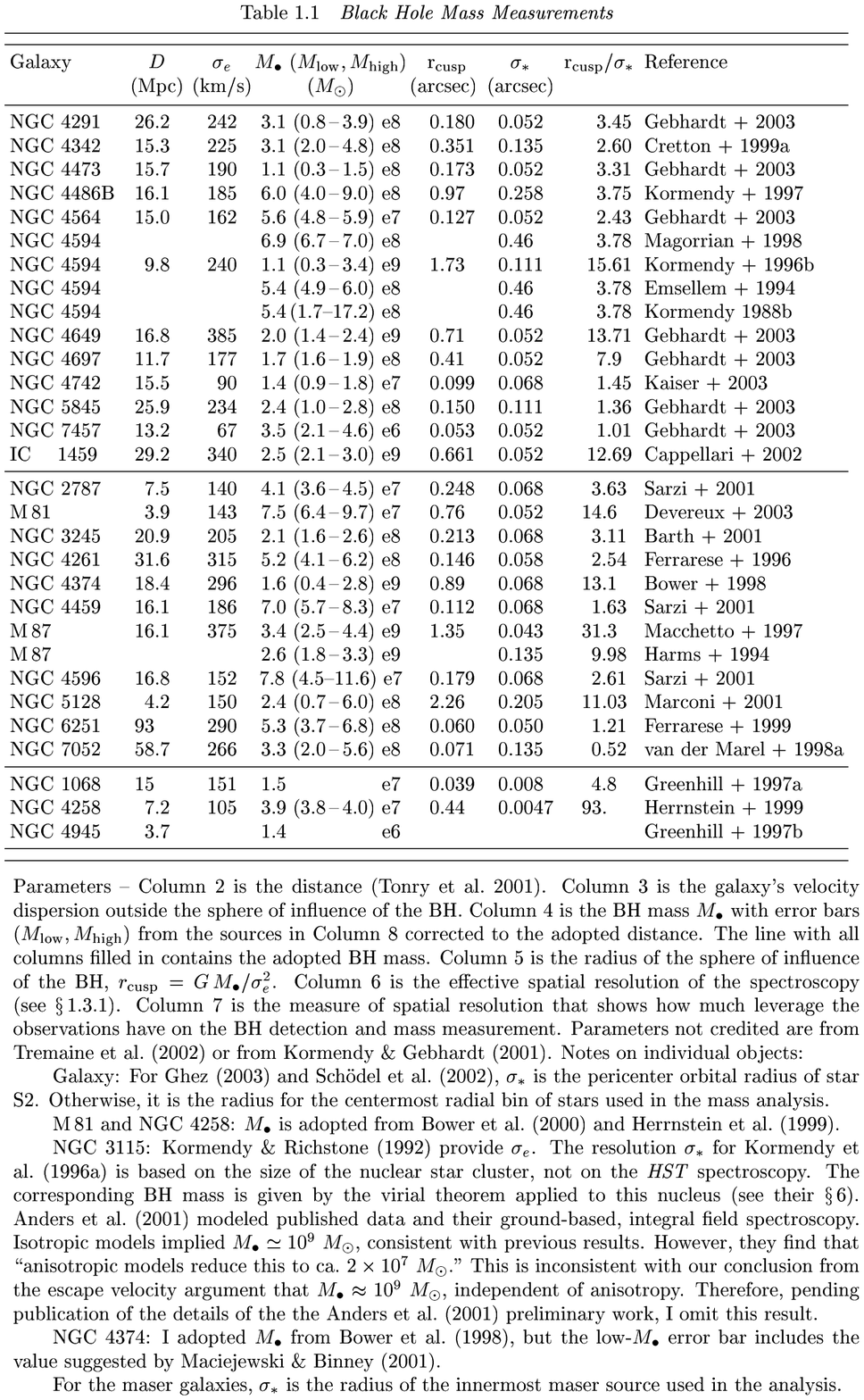}

\eject

      The stellar-dynamical BH search began with two papers on M{\ts}87
by Young \etal (1978) and by Sargent \etal (1978).  Based on the
non-isothermal (cuspy) surface brightness profile of its core and an
observed rise in velocity dispersion toward the center, they showed
that M{\ts}87 contains an $M_\bullet$ $\simeq 4 \times
10^9$ $M_\odot$ MDO if the stellar velocity distribution is isotropic.
At about the same time, it became clear that almost no giant
ellipticals like M{\ts}87 are isotropic (e.g., Illingworth 1977;
Binney 1978) and that anisotropic models can explain the cuspy core and
the dispersion gradient without a BH (Duncan \& Wheeler 1980; Binney \&
Mamon 1982; Richstone \& Tremaine 1985; Dressler \& Richstone 1990).
Nevertheless, the Young and Sargent papers were seminal.  They
set the field in motion.

      The dynamical detection of dark objects in galaxy centers began
with the discovery of an $M_\bullet \approx 10^{6.5}$ $M_\odot$~mass~in~M{\ts}32
(Tonry 1984, 1987; Dressler \& Richstone 1988), a $10^{7.5}$ $M_\odot$ object in
M{\ts}31 (Dressler \& Richstone 1988; Kormendy 1988a), and $10^{9}$ $M_\odot$
objects in NGC 4594 (Kormendy 1988b) and NGC 3115 (Kormendy
\& Richstone 1992).  The observations were ground-based with resolution FWHM
$\approx$\ts1$^{\prime\prime}$.  The BH case in our Galaxy developed slowly (see
Genzel, Hollenbach, \& Townes 1994; Kormendy \& Richstone 1995 
for reviews), for two reasons.  Dust extinction made
it necessary to use infrared techniques that were just being developed in the
early 1990s.  And the $M_\bullet$ measurement in our Galaxy requires the study
of a relatively small number of stars that are bright enough to be observed
individually.  As a result, graininess in the light and velocity distributions
becomes a problem.  On the other hand, the Galactic Center is very close, so
progress in the past decade has been spectacular.  Now the Galaxy is by
far the best supermassive BH case (\S\thinspace1.3.2).

      The BH search speeded up dramatically once {\it HST\/} provided
spatial resolution a factor of 3 to 10 better than ground-based
telescopes (see Kormendy \& Gebhardt 2001 for a review).  By now,
almost all galaxies in which BHs were discovered from the ground have
undergone several iterations of improved spatial resolution.  Analysis 
machinery has improved just as dramatically.  This is an opportune time
to take stock of the past 15 years of progress.   Are the detections of 
central dark objects reliable?  Are the derived masses robust?  And are 
the dark objects really BHs?  The BH search is starting to look like a 
solved problem; assuming this, emphasis has shifted to demographic studies
of BHs and their relation to galaxy evolution (see Richstone et al.~1998; 
Ho 1999; Kormendy \& Gebhardt 2001; Richstone 2003 for reviews).  Is this 
a reasonable attitude?  Sanity checks are the purpose of this paper.

\section{The History of BH Mass Measurements}

      The history of supermassive BH mass measurements is summarized in Table 
1.1.  In focusing on this history, I will be concerned with whether we achieve
approximately the accuracies that we believe.  That is, I concentrate on errors
of \gapprox\ts0.2 dex.  To what extent hard work can further squeeze the
measurement errors is discussed by Gebhardt (2003).  

       In Table 1.1, horizontal lines separate BH detections based on
stellar dynamics (first group), ionized gas dynamics (middle), and maser
dynamics (last group).  All multiple stellar-dynamical $M_\bullet$ estimates
for the same galaxy are listed.  Our Galaxy, M{\ts}31, M{\ts}32, NGC 3115, 
NGC 3377, and NGC 4594 have all been measured by at least two competing groups.
M{\ts}81 has been observed independently in stars and ionized gas; both
measurements are listed and they agree.  
% M{\ts}87 has ionized gas $V(r)$ measurements from two {\it HST\/} 
% instruments; these results agree also.  
However, consistency checks of $M_\bullet$ values based on ionized
gas dynamics have revealed some problems in other galaxies; these are discussed
by Maciejewski \& Binney (2001), Barth et al.~(2001), Verdoes Kleijn et
al.~(2002), Barth (2003), and Sarzi (2003).  I have not included
all multiple measurements based on ionized gas dynamics. 

% XXX Axon didn't turn in paper

%\vfill\eject

\section{How Robust Are Stellar-Dynamical BH Mass Estimates?}

\subsection{The History of the BH Search As Seen Through Work on M{\ts}32}

\pretolerance=10000  \tolerance=10000

      M{\ts}32 was the first application of many improvements in spatial
resolution, in kinematic analysis techniques, and in dynamical modeling machinery.   It provides an excellent case study for a review of these
developments.  Figure 1.1 illustrates the remarkable result that BH mass
estimates for M{\ts}32 have remained stable for more than 15 years while a
variety of competing groups have improved the observations and analysis\footnote{The referee suggests that this result is caused by two effects
that accidentally cancel because spatial resolution and dynamical models have
improved in parallel.  He suggests (1) that $M_\bullet$ estimates increase with
improving spatial resolution because we reach farther into the BH sphere of
influence and (2) that $M_\bullet$ estimates decrease as dynamical models get
more sophisticated because the models have more freedom to tinker the orbital
structure to fit the data without a BH.  I disagree.  (1) Reaching farther into
the BH sphere of influence should not change $M_\bullet$ if we model the stellar
dynamics adequately well.  Instead, we should get more "leverage" and smaller
mass error bars.  Of course, if we model the physics incorrectly, then more
leverage may result in a systematic change in $M_\bullet$. But the change could
go either way, depending on how the models err in approximating the true
velocity anisotropy.  In fact, Figure 1.4  shows that improving the spatial
resolution does not increase the $M_\bullet$ values given by the Gebhardt et
al.~(2003) three-integral models, although it does, as expected, improve the
error bars.  For the Magorrian et al.~(1998) models, improving the resolution
decreases $M_\bullet$, an effect opposite to that predicted by the referee. 
(2) Improving modeling techniques
provides more degrees of freedom on the orbital structure, but modeling
programs do not have any built-in desire to decrease the BH mass.  Instead, 
they have instructions to fit the data.  Again, if the real orbital structure
is sufficiently well approximated by simple models, then making the models more
complicated will not change the BH mass.  And if the orbital structure is not
well approximated by the simple models, then better models could just as easily
increase $M_\bullet$ as decrease it.  However, the low-mass error bar on
$M_\bullet$ will decrease, for the reason the referee suggests.  The high-mass
error bar will increase.  As a result, the error bars become larger and more
realistic.  This effect is evident in Table 1.1.  I conclude that the
consistency of $M_\bullet$ estimates in Figures 1.1 and 1.2 tells us
something important, namely that we have been modeling the stellar dynamics of
power-law galaxies well enough to derive robust BH masses.}.

      The BH in M\ts32 was discovered as early as possible, when the spatial
resolution was so poor that $r_{\rm cusp}/\sigma_*$ $<$ 1.  This is
not surprising, given the importance of the~problem.  In astronomy as
in other sciences, if you wait for a 5\ts$\sigma$ result, someone else is
likely to make the discovery when it is still a 2\ts$\sigma$ result. The trick
is to be careful enough to get the right answer even when the result is
uncertain.  Tonry (1984, 1987) got within a factor of 2.5 of the current
best BH mass even though he made serious simplifying~assumptions.
%He used adjustable analytic approximations to the rotation curve; this is not
%a problem.  
His spectra did not resolve the intrinsic velocity dispersion
gradient near the center; rotational line broadening accounted for the apparent
dispersion gradient.  Without an intrinsic dispersion gradient, his models were
guaranteed not to be self-consistent, because there was no dynamical support 
in the axial direction.  Despite this approximation, Tonry derived $M_\bullet
\simeq (6~\rm to~8) \times 10^6$ $M_\odot$, close to the modern value.  Poor
spatial resolution allowed considerable freedom to interpret dispersion
gradients as unresolved rotation; since $V$ and $\sigma$ contribute comparably
to the dynamical support, trading one for the other results in no large change
in $M_\bullet$.

	 The spatial resolution of the spectroscopy improved by a factor of 30
from the discovery observations (Tonry 1984) to the Space Telescope Imaging
Spectrograph (STIS) data from 
{\it HST\/}.  In Column 6 of Table~1.1, the Gaussian dispersion radius
of the PSF is estimated as follows.  First, I estimate the resolution in the
directions parallel and perpendicular to the slit as
$\sigma_{*\parallel}$, the sum in quadrature of the radius $\sigma_{*\rm
tel}$ of the telescope PSF and of 1/2 pixel, and $\sigma_{*\bot}$,
the sum in quadrature of the radius of the telescope PSF and half of
the slit width.  The {\it HST\/} PSF was modeled in van der Marel, de Zeeuw, \&
Rix (1997b) as the sum of three Gaussians; for all {\it HST\/} observations, I
use $\sigma_{*\rm tel} \simeq 0\sd036$, the best single
Gaussian dispersion radius that fits this sum.  Finally, the
effective $\sigma_*$ is the geometric mean of $\sigma_{*\parallel}$ and
$\sigma_{*\bot}$.  I do not take into account slit centering
errors; for some observations, these are larger than $\sigma_*$.

\begin{figure*}[ht!]
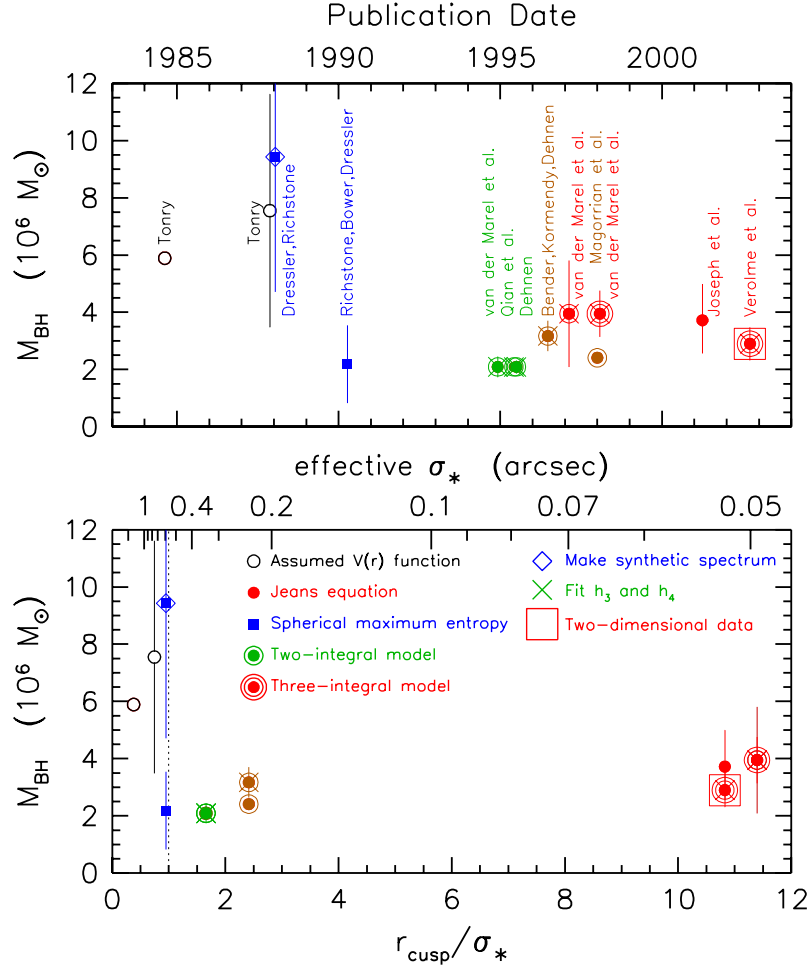
 
\centering
\vspace{12.72cm}
\bw{ \includegraphics{m32mbh_carnegie.ps}}
\col{\includegraphics{m32mbh_carnegie.cps}}
\vskip 0pt
\caption{History of the stellar-dynamical BH search as seen through work
         on~M{\ts}32: derived BH mass as a function of (top)  publication date
         and (bottom) spatial resolution.  Resolution is 
         measured along the top axis by the Gaussian dispersion radius
         $\sigma_*$ of the effective PSF (see text).  More relevant physically
         (bottom axis) is the ratio of the radius of the sphere of influence 
         of the BH, $r_{\rm cusp} = GM_{\bullet}/\sigma^2$, to $\sigma_*$.
         If $r_{\rm cusp}/\sigma_*$ \lapprox \ts1, then the measurements are
         dominated by the mass distribution of the stars rather than by the BH.
         If $r_{\rm cusp}/\sigma_* \gg 1$, then we
         reach well into the part of the galaxy where velocities are dominated
         by the BH.  Symbols shapes encode improvements in
         observations or kinematic measurements (right key) and in dynamical
         modeling techniques (left key).  The data are listed in Table 1.1.
\label{fig1}}
\end{figure*}

  %  For example, for the Macchetto et al.~(1997) observation of M{\ts}87, 
  %  $\sigma_* = 0\sd043$ while the slit missed the galaxy center by 0\sd07.

      Dressler \& Richstone (1988) and Richstone, Bower, \& Dressler (1990)
followed with better observations and analysis.  They fitted spherical maximum entropy models including velocity anisotropy.  By this time, it was well known
that unknown velocity dispersion anisotropy was the biggest uncertainty in 
$M_\bullet$ measurements based on stellar dynamics.  They were unable to explain
the central kinematic gradients in M\ts32 without a BH.  Rapid confirmation of
Tonry's BH detection contributed to the early acceptance of this subject.

%\eject

      Since then, dynamical modeling machinery has improved remarkably.
The next major step defined the state of the art from 1995 through 1997.
This was the use of two-integral models that included flattening
and velocity dispersion anisotropy.  Essentially simultaneous work by
van der Marel et al.~(1994b), Qian et al.~(1995), and Dehnen (1995) all derived
$M_\bullet = 2.1 \times 10^6$ $M_\odot$ from van der Marel's data.  Soon 
thereafter, Bender, Kormendy, \& Dehnen (1996) got $3.2 \times 10^6$ $M_\odot$
using the same machinery on CFHT data of \hbox{slightly higher} resolution.  
The limitation of these models, as the authors realized,
was the fact that two-integral models are approximations.  They work best for
cuspy and relatively rapidly rotating galaxies like M{\ts}32, but they are
not fully general.  Still, by this time, it was routine to measure not just the
first two moments of the line-of-sight velocity distributions (LOSVDs) --- that
is, $V$ and $\sigma$ --- but also the next two coefficients $h_3$ and $h_4$ in a
Gauss-Hermite expansion of the LOSVDs.  These measure asymmetric and
symmetric departures from Gaussian line profiles.  In a transparent galaxy that
rotates differentially, projection guarantees that $h_3 \ne 0$.  In general,
$h_3$ is antisymmetric with $V$.  A galaxy containing a BH is likely to have
$h_4 > 0$; that is, an LOSVD that is more centrally peaked than a Gaussian.  The
reason is that stars close to the BH move very rapidly and give the LOSVD
broader symmetric wings than they would otherwise have (van der Marel 1994).
Thus, as emphasized especially by van der Marel et al.~(1994a), measuring and
fitting $h_3$ and $h_4$ adds important new constraints both to the stellar
distribution function and to the BH detection and mass determination.

% Luis: "adds", not "add".

      {\it HST\/} Faint Object Spectrograph (FOS) observations of M\ts32 were
obtained by van der Marel et al.~(1998b).  These authors further ``raised the bar'' on BH mass measurements 
by fitting their data with three-integral dynamical models constructed using
Schwarzschild's (1979) method.  Such models now define the state of the art
(see Cretton et al.~1999b; Gebhardt et al.~2000a, 2003; Richstone et 
al.~2003 for more detail).

\def\bck{\kern -2pt} 

      Finally, the most thorough data set and modeling analysis for M\ts32 is
provided by Verolme et al.~(2002).  They use the SAURON two-dimensional
spectrograph to measure $V$, $\sigma$, $h_3$, and $h_4$ in the central
9$^{\prime\prime}$ \bck$\times$ \bck11$^{\prime\prime}$.  Also, {\it HST\/} STIS
spectroscopy (Joseph et al.~2001) provides improved data near the BH.  These
observations fitted with three-integral models for the first time break the
near-degeneracy between the stellar mass-to-light ratio, $M/L$, and the
unknown inclination of the galaxy.  Because the mass in stars is better
known, the BH mass is more reliable.  Again, the derived BH mass is similar to
that given in previous analyses, $M_\bullet = (2.9 \pm 0.6) \times \bck10^6$
$M_\odot$.  

      So the BH mass derived for M\ts32 has remained almost unchanged while the
\hbox{observations} and analysis have improved dramatically.~It was exceedingly
important to our confidence in the BH detection to test whether the apparent 
kinematic gradients near the center could be explained without a BH.  Asked to
do this, a dynamical modeling code attempts to fine-tune the stellar velocity
dispersion anisotropy.  In general, it tries to add more radial orbits near the
center, because doing so implies less mass for the same $\sigma$.  Nowadays,
its freedom to tinker is severely restricted by the need to match the full
LOSVDs.  However, even simple approximations to the dynamical structure gave 
essentially the correct BH mass.  {\it That is, M{\ts}32 does not use its
freedom to indulge in perverse orbit structure.}  The following sections
show that this is also true in our Galaxy, M\ts31, NGC 3117, NGC 3377, and
NGC 4594.  Dynamical mass modeling is relatively benign in such galaxies
that have power-law profiles (for more details, see 
Kormendy et al.~1994; Lauer et al.~1995; Gebhardt et al.~1996; Faber et 
al.~1997; Lauer 2003).  It would not be safe to assume that this result
applies equally well to galaxies with cuspy cores.

\subsection{The Best Case of a Supermassive Black Hole: Our Galaxy}

      Figure 1.2 summarizes the history of BH mass measurements in galaxies 
with observations or stellar-dynamical mass analyses by different research
groups.  The BH case that has improved the most is the one in our Galaxy.  
Both the evidence for a central dark object and the arguments that this is a 
BH and not something less exotic like a cluster of dark stars are better in 
our Galaxy than anywhere else.

\begin{figure*}[hb!]
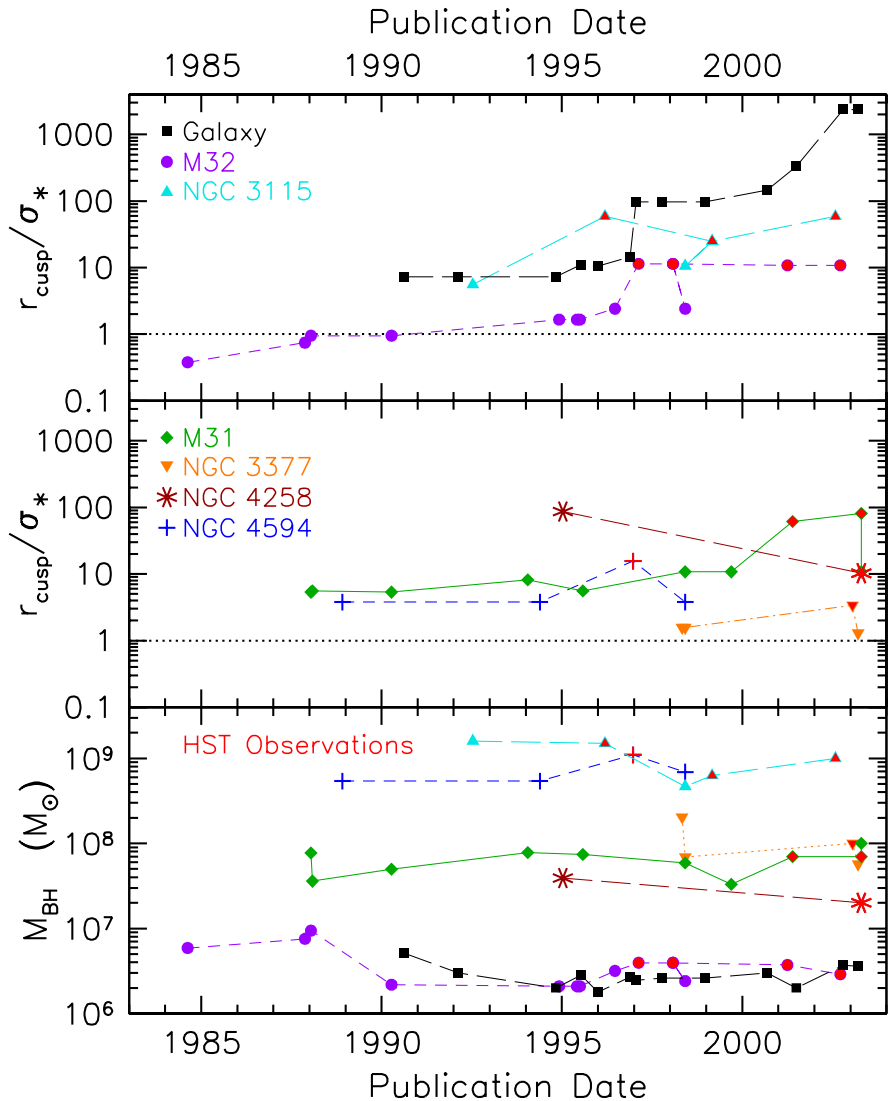

\centering
\vspace{14.44cm}
\bw{ \includegraphics{bh_discovery_carnegie_3panels.ps}}
\col{\includegraphics{bh_discovery_carnegie_3panels.cps}}
\vskip 0pt
\caption{Effective resolution of the best spectroscopy (top two panels) and
         resulting BH mass estimates (bottom) versus
         publication date.  The data are listed in Table 1.1. For M{\ts}31 and
          M{\ts}32, steep rises in $r_{\rm cusp}/\sigma_*$ occur when
         {\it HST\/} was first used to observe the galaxies.  For our Galaxy,
         two jumps in $r_{\rm cusp}/\sigma_*$ occur when the kinematic
         work switched from radial velocities to proper motions in
         the Sgr A* star cluster and when the first nearly complete
         stellar orbit in that cluster was observed.}
\label{fig2}
\end{figure*}

\clearpage % Note that the backlog of floats is cleared after this command.

% FloatBarrier

      A complete review of the BH search in our Galaxy is beyond the scope
of this paper.  Early work is discussed in Genzel \& Townes (1987);
Genzel et al.~(1994); Kormendy \& Richstone (1995), and in conference
proceedings such as Backer (1987), Morris (1989), and Genzel \& Harris (1994).
Observations of our Galactic Center benefit from the fact that it is 100 times
closer than the next nearest good BH cases, M\ts31 and M\ts32.  For a distance
of 8 kpc, the scale is 25\sd8 pc$^{-1}$.   Early gas- and stellar-dynamical
studies suggested the presence of a several-million-solar-mass dark object. 
In Table 1.1 and Figure 1.2, I date the convincing case for a BH to Sellgren
et al.~(1990) and to Kent (1992). Since then, two dramatic improvements in
spatial resolution have taken place.

      Research groups led by Reinhard Genzel and Andrea Ghez have pioneered the
use of speckle interferometry and, more recently, adaptive optics imaging and
spectroscopy to achieve spatial resolutions good enough to resolve a tiny
cluster of stars (radius $\sim 1^{\prime\prime}$) that surrounds the compact
radio source Sgr A* at the Galactic Center.  The Sgr A* cluster is so tiny
that stars move fast enough to allow us to observe proper motions.  This 
provides a direct measure of the velocity dispersion anisotropy.  It is not
large.   The derived central mass is about $2.5 \times 10^6$ $M_\odot$.  And, 
even though the number density of stars is higher than we observe anywhere
else, the volume is so small that the stellar mass is negligible.  The advent 
of proper motion measurements accounts for the jump in $r_{\rm cusp}/\sigma_*$
at the start of 1997.

      A second jump in $r_{\rm cusp}/\sigma_*$ has just occurred as a result
of an even more remarkable observational coup.  As reviewed in this volume
by Ghez (2003), Sch\"odel et al.~(2002), Ghez (2003), and Ghez et
al.~(2003) have independently measured several individual stellar orbits through
pericenter passage.  In the case of star S2, more than half of an orbit has been
observed (period = 15.78 $\pm$ 0.82 years).  The orbit is closed, so the
controlling
mass resides inside $r_{\rm peri} \simeq 0\sd0159 \simeq 0.00062$ pc $\simeq$
127 AU $\simeq$ 1790 Schwarzschild radii.  This accounts for the current jump 
in spatial resolution.  As measurement accuracies improve, the observation of
individual closed orbits will rapidly obsolete the complicated analysis of
stellar distribution functions that describe ensembles of stars at larger radii.
Rather, the analysis will acquire the much greater rigor inherent in the
two-body problem.  Arguably the orbit of S2 already contributes as much to our
confidence in the BH detection as all stars at larger radii combined.
The best-fitting BH mass, $M_\bullet =(3.7 \pm 0.4) \times 10^6$ $M_\odot$, is
in good agreement with, but slightly larger than, the value derived from the
stellar-dynamical modeling.  This leads to an important point: The
above comparison in our Galaxy and a similar one in NGC 4258 (see the next
section) are currently the only reliable external checks on our 
stellar-dynamical modeling machinery.  The measurement accuracies are not good 
enough yet to show whether the models achieve the accuracies that we expect
for the best data ($\pm$30\%: Gebhardt 2003).  But neither test points to
modeling errors that range over a factor of $\sim 6$ as feared by Valluri,
Merritt, \& Emsellem (2003).   

      Finally, these new observations have an implication that is
actually more fundamental than the mass measurement.  They restrict the dark
mass to live inside such a small radius that even neutrino balls 
(Tsiklauri \& Viollier 1998, 1999; Munyaneza, Tsiklauri, \& Viollier 1998, 
1999; Munyaneza \& Viollier 2002) 
with astrophysically allowable neutrino masses are excluded.  The exclusion
principle forces them to be too fluffy to be consistent with the radius
constraints.  Dark clusters of brown dwarf stars or stellar remnants were
already excluded (Maoz 1995, 1998) --- brown dwarfs would collide, merge,
and become visible stars, and stellar remnants would evaporate via 
relaxation processes.  The maximum lifetime of dark cluster alternatives to 
a BH is now a few times $10^5$ yr (Sch\"odel et al.~2002).

%      Our Galaxy therefore addresses our confidence in the BH paradigm in
%two fundamental ways.  Very large improvements in spatial resolution and
%analysis machinery again confirm tha the derived BH mass is robust and
%insensitive to quirky orbital structure.  This supports our confidence that
%dynamical analyses are also trustworthy in galaxies where the checks 
%inherent in Figure 1.3 are not yet possible.  And the astrophysical issue
%of whether we are indeed detecting a BH or ``just'' a cluster of dark stars ---
%which would, in any case, be exceedingly exotic --- is largely solved.

\subsection{The Best Test of Stellar Dynamical $M_\bullet$ Estimates: NGC 4258}

      The galaxy that stands out as having the most reliable BH mass measurement
is NGC 4258.  Very Long Baseline Array measurements of its nuclear water maser
disk reach to within 0\sd0047 = 0.16 pc of the BH (Miyoshi et al.~1995).~The
rotation curve, \hbox{$V(r) = 2180~(r/0\sd001)^{-1/2}$} km s$^{-1}$, is Keplerian
to high precision.  Proper motion and acceleration observations of the masers in
front of the Seyfert nucleus are consistent with the radial velocity
measurements along the orbital tangent points (Herrnstein et al.~1999).  All
indications are that the rotation is circular.  Therefore \hbox{$M_\bullet =
(3.9 \pm 0.1) \times 10^7$ $M_\odot$} is generally regarded as bomb-proof. 

      This provides a unique opportunity to test the three-integral dynamical
modeling machinery used by the Nuker team (Gebhardt et al.~2000a, b; 2003;
Richstone et al.~2003).  NGC 4258 contains a normal bulge much like the one in
M{\ts}31 (Kormendy et al.~2003a). Siopis et al.~(2003) have obtained {\it HST\/}
STIS spectra and WFPC2 images of NGC 4258.  The STIS spectroscopy has spatial
resolution $r_{\rm cusp}/\sigma_* \simeq 8.4$ well within the range of the BH
discoveries in Table 1.1.  The kinematic gradients are steep, consistent with
the presence of a BH.  Three-integral models are being calculated as I write
this; the preliminary result is that $M_\bullet = (2 \pm 1) \times 10^7$
$M_\odot$.  The agreement with the maser $M_\bullet$ is fair.  The problem is
the brightness profile, which involves more complications than in most BH 
galaxies.  A color gradient near the center may be a sign of dust obscuration,
and correction for the bright AGN (Chary et al.~2000) is nontrivial.  Both
problems get magnified by deprojection.

\subsection{A Case History of Improving Spatial Resolution: NGC 3115}

      One sanity check on BH detections is that apparent kinematic gradients
should get steeper as the spectroscopic resolution improves.  We have
seen this test work in M{\ts}32 and in our Galaxy.  This section is a brief
discussion of NGC 3115.  At $r_{\rm cusp}/\sigma_* = 59$, NGC 3115 is surpassed
in spectroscopic resolution only by our Galaxy, NGC 4258, and M{\ts}31.

      Exploiting the good seeing on Mauna Kea, Kormendy \& Richstone (1992)
found a central dark object of $10^9$ $M_\odot$ in NGC 3115 using the
Canada-France-Hawaii Telescope (CFHT).  The resolution was not marginal; 
$r_{\rm cusp}/\sigma_* \simeq 5.5$.  This is higher than the median for
{\it HST\/} BH discoveries in Figure 1.3 (\S\ts1.3.7).  Since then, there have
been two iterations in improved spectroscopic resolution (Kormendy et 
al.~1996a).  The apparent central velocity dispersion increased correspondingly:
it was $\sigma = 295 \pm 9$ km s$^{-1}$ at $r_{\rm cusp}/\sigma_* \simeq 5.5$,
$\sigma = 343 \pm 19$ km s$^{-1}$ at $r_{\rm cusp}/\sigma_* \simeq 10.6$
(CFHT plus Subarcsecond Imaging Spectrograph), and 
$\sigma = 443 \pm 18$ km s$^{-1}$ at $r_{\rm cusp}/\sigma_* \simeq 59$
({\it HST\/} FOS).  These are projected velocity
dispersions: they include the contribution of foreground and
background stars that are far from the BH and so have relatively small
velocity dispersions.  However, NGC 3115 has a tiny nuclear star cluster
that is very distinct from the rest of the bulge.  It is just the sort of
high-density concentration of stars that we always expected to find around 
a BH.  From a practical
point of view, it is a great convenience, because it is easy to subtract
the foreground and background light as estimated from the spectra 
immediately adjacent to the nucleus.  This procedure is analogous to sky
subtraction.  It provides the velocity dispersion
of the nuclear cluster by itself and is, in effect, another way to increase
the spatial resolution.  The result is that the nuclear cluster has a
velocity dispersion of $\sigma = 600 \pm 37$ km s$^{-1}$.  The effective 
spatial resolution of this measurement is not determined by the spectrograph
but rather by the half-radius $r_h = 0\sd052 \pm 0\sd010$ of the nuclear
cluster.  This is smaller than the entrance aperture of the FOS. It implies
that $r_{\rm cusp}/\sigma_* \simeq 59$, as quoted in Table 1.1.

      The nucleus allows us to estimate the BH mass
independent of any velocity anisotropy.  If the nucleus consisted only of old
stars with the mass-to-light ratio measured for the bulge, then its mass would
be $\sim 4 \times 10^7$ $M_\odot$ and
its escape velocity would be $\sim$~352 km s$^{-1}$.  This is much smaller than
the observed velocities of the stars.  The nucleus would fly apart in a few
crossing times $T_{\rm cross}$.  But $T_{\rm cross} \simeq 16,000$ yr is
very short.  Therefore, a dark object of $10^9$\ts$M_\odot$ must be
present to confine the stars within the nucleus.

\subsection{A Comparison of Ground-Based and {\it HST\/} Studies of NGC 3377 and NGC 4594}

       Besides M{\ts}31 (\S\ts1.4), {\it HST\/} has confirmed ground-based BH
detections in two more galaxies (Fig. 1.2). 

       NGC 4594, the Sombrero galaxy, was observed with the CFHT by Kormendy
(1988b), yielding a BH mass of $M_\bullet \approx 10^{8.7}$ $M_\odot$.  Resolution
was average for BH detections; \hbox{$r_{\rm cusp}/\sigma_* = 3.8$.} The galaxy
was reobserved with {\it HST\/} by Kormendy et al.~(1996b) using the FOS at $r_{\rm cusp}/\sigma_* \approx 15.6$.  They confirmed the BH detection and quoted a
slightly higher mass of $10^9$ $M_\odot$.  This test is weaker than those
quoted above because the same research group was involved and because
three-integral models were not constructed.  However, independent dynamical
models by Emsellem et al.~(1994) agree very well with the results
in Kormendy (1988b).

      NGC 3377 also has a CFHT BH detection; $r_{\rm cusp}/\sigma_* = 1.57$
(Kormendy et al.~1998).  The BH mass was $M_\bullet = (2 \pm 1)
\times 10^8$ $M_\odot$.  Gebhardt et al.~(2003) reobserved the galaxy with {\it HST\/}
at $r_{\rm cusp}/\sigma_* = 3.4$.  The improvement in resolution is smaller than normal because the CFHT seeing was very good and because the {\it HST\/} FOS aperture
size was 0\sd2.  Nevertheless, the improvement is substantial.  Also, the
analysis machinery was updated; Kormendy et al.~(1998) fitted analytic
approximations to $V$ and $\sigma$ and, independently, spherical maximum
entropy models with post-hoc flattening corrections.  Gebhardt et al.~(2003)
fitted three-integral models.  They obtained $M_\bullet = 1.0_{-0.1}^{+0.9}
\times 10^8$ $M_\odot$, confirming the earlier result.  Also, 
Cretton et al.~(2003) report two-dimensional spectroscopy in the
inner $6^{\prime\prime}$ \bck$\times$ \bck$3^{\prime\prime}$ of NGC 3377.  Three-integral models give $M_\bullet = 5.7_{-2.3}^{+5.6} \times 10^7$
$M_\odot$, corrected to our adopted distance.  Again, the published results are
consistent.

\subsection{Robustness of Stellar-Dynamical $M_\bullet$ Values. I.
            Conclusion from \S\S\ts1.3.1 -- 1.3.5}

      All of the ground-based, stellar-dynamical BH detections 
discussed in Kormendy \& Richstone (1995) have now been confirmed at higher
spatial resolution and with more sophisticated modeling machinery.  All of the 
original mass estimates agree with the best current values to factors of 
2 -- 3 or better.

     Given the above tests, given the agreement between the BH
parameter correlations implied by the dynamics of stars, of
ionized gas, and of maser gas, and especially given the tightness of
the scatter in the $M_\bullet$ -- $\sigma$ correlation, it seems
unlikely that $M_\bullet$ values are still uncertain to factors of
several, as suggested by Valluri et al.~(2003).  Nevertheless, so much is at
stake that we must continue to test the stellar-dynamical modeling codes.  
For example, triaxiality is not yet included.  It is unlikely that
triaxiality provides enough new degrees of freedom to greatly change the
results; very triaxial configurations would have been seen with {\it HST\/}.
But checking the consequences of triaxiality is under way by the SAURON team.

      All papers contain simplifying assumptions.  Science is the art of
getting the right answer using approximate analysis of imperfect data.  We
should not get complacent, but we appear to be doing reasonably well.

\subsection{Application: {\it HST\/} BH Discoveries}

      Having shown from repeat observations at better spatial resolution how
well we do when $r_{\rm cusp}/\sigma_* \simeq 1$ -- 10, we now apply these
results to {\it HST\/} BH discoveries that do not have repeat measurements.

\begin{figure*}[ht!]
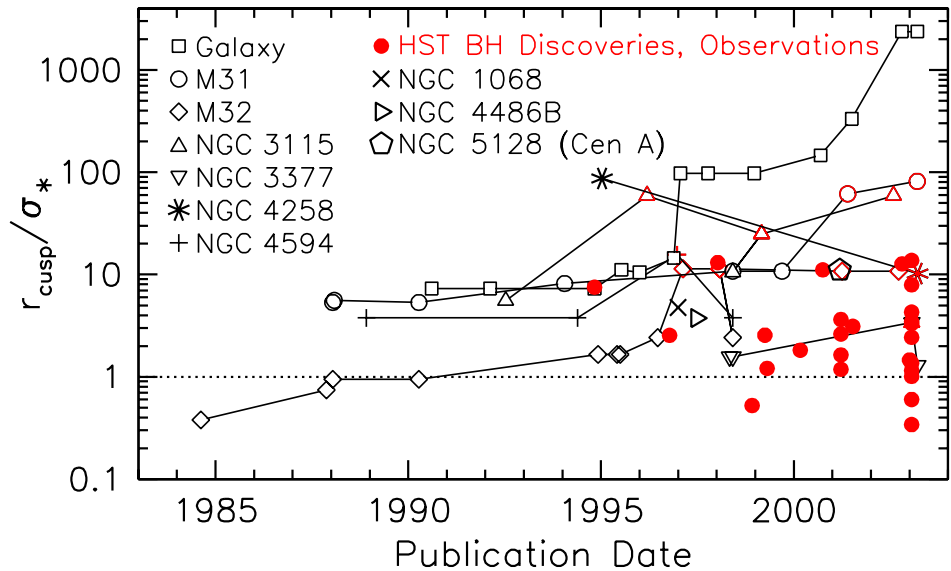
 % [hb!]
\centering
\vspace{7.5cm}
\bw{ \includegraphics{bh_discovery_carnegie_toppanel.ps}}
\col{\includegraphics{bh_discovery_carnegie_toppanel_red.cps}}
\vskip 0pt
\caption{Spectroscopic spatial resolution for all BH discoveries in Table 1.1.
         Galaxies with repeat measurements are from Figures 1.1 and 1.2. 
         %{\it HST\/} observations are in red. 
         Note that {\it HST\/} and \hbox{ground-based} BH discoveries
         have similar distributions of $r_{\rm cusp}/\sigma_*$.
         However, {\it HST\/} has 5 -- 10 times better spatial resolution in arcsec
         than ground-based observations (absent adaptive optics).  This means
         that {\it HST\/} is being used to discover lower-mass BHs in more distant
         galaxies.}
\label{fig1.4}
\end{figure*}

%\clearpage % Note that the backlog of floats is cleared after this command.

      Figure 1.3 shows the distribution of $r_{\rm cusp}/\sigma_*$ values for
all BH detections made with {\it HST\/}.  It contains a number of surprises.  Contrary
to popular belief, {\it {\it HST\/} BH discoveries are not being made with much better
spatial resolution than ground-based BH discoveries.\/}  Only a few of the best
{\it HST\/} cases have $r_{\rm cusp}/\sigma_* \simeq 10$ comparable to the ground-based
BH detections in our Galaxy, in M\ts31, and in NGC 3115. On average, {\it HST\/} BH
discoveries are being made at lower $r_{\rm cusp}/\sigma_*$ values than those
made from the ground.  Several have $r_{\rm cusp}/\sigma_* 
< 1$, similar to the early measurements of M\ts32.   I am not
suggesting that {\it HST\/} and ground-based spatial resolutions are similar
{\it in arcsec\/}.  {\it HST\/} is better by a factor of 10 (if a 0\sd1
slit is used) or at least 5 (for measurements with the 0\sd2 aperture or slit).
What is really going on is this: The ground-based observations ``used up'' the
best galaxies.  For example, our Galaxy, M\ts31, and M{\ts}32 are unusually
close, and NGC 3115 has an unusually large BH mass fraction.  So {\it HST\/} is
necessarily being used on more distant galaxies or ones that have smaller BH
mass fractions.  This puts the exceedingly important contributions of {\it HST\/} into perspective:

      (1) {\it HST\/} did not find the strongest BH cases.  NGC 4258
and our Galaxy were observed from the ground.  {\it HST\/} observations of
NGC 4258 serve to test the stellar-dynamical models.

      (2) {\it HST\/} has confirmed and greatly strengthened the BH cases for
BH discoveries made from the ground.  The spectroscopic resolutions 
$r_{\rm cusp}/\sigma_*$ for M{\ts}31 and for NGC 3115 are now essentially as
good as that for the famous maser case, NGC 4258.

      (3) {\it HST\/} did not revolutionize BH detections by finding them at higher resolution.

      (4) {\it HST\/} has revolutionized the BH search by allowing us
to find smaller BHs and ones in more distant galaxies.  This has two
important implications.

      (5a) There has always been a danger that ground-based observations would
be biased in favor of BHs that are unusually massive. Any such bias is rapidly
being diluted away.  In fact, it was not large.  Kormendy \& Richstone (1995) 
found from ground-based observations that the mean ratio of BH mass to bulge
mass was 
$\langle M_{\bullet}/M_{\rm bulge}\rangle$\ts$=$\ts$0.0022^{+0.0016}_{-0.0009}$
(they averaged log\ts$M_{\bullet}/M_{\rm bulge}$ for eight BH detections, six 
made with stellar dynamics and one each with masers and ionized gas disks).   
Now, the data in Table 1.1
give $\langle M_{\bullet}/M_{\rm bulge}\rangle$\ts$=$\ts$0.0013$ 
(Kormendy \& Gebhardt 2001; Merritt \& Ferrarese 2001).  

      (5b) {\it HST\/} has made it possible to detect canonical BHs (ones within the
scatter of the $M_\bullet$ correlations) out to the distance of the Virgo
cluster.  The largest BHs can be detected several times farther away.  This has
revolutionized the subject of BH demographics.  We now have enough detections
to address the question of how BH growth is related to galaxy formation.

     (6) As Figure 1.3 emphasizes, this subject has speeded up enormously 
because of {\it HST\/}.

\subsection{Caveat: Cuspy Core Galaxies}

      The caveat to this rosy story is that the above tests were carried out
for galaxies with ``power-law profiles'' (Lauer et al.~1995).
The physical distinction between such galaxies and ones with cuspy cores is
discussed by Kormendy et al.~(1994), Lauer et al.~(1995), Gebhardt et
al.~(1996), and especially Faber et al.~(1997).  The observations
imply that cuspy core galaxies have more anisotropic velocity distributions
than do power-law galaxies (Kormendy \& Bender 1996).  They are fundamentally
more difficult for the BH search (Kormendy 1993).  The shallower volume
brightness profile $\rho(r)$ gives, in projection, less luminosity weight to the
stars in the sphere of influence of the BH.  The $d\ts{\ln{\rho}}/d\ts{\ln{r}}$
term in the mass derivation is smaller and more easily cancelled by the effects
of velocity anisotropy, which is larger than in power-law galaxies.  Stellar
dynamical BH detections in cuspy core galaxies are few and not well tested.  
Comparisons between stellar-dynamical and gas-dynamical $M_\bullet$ measurements
do not show universally good agreement.  BH masses in cuspy core galaxies are
more uncertain than those in power-law galaxies, and the above conclusions 
cannot confidently be applied to them.  We need better tests of BH detections
in core galaxies.

\subsection{Robustness of Stellar-Dynamical $M_\bullet$ Values. II.
            What Resolution Do We Need?}

      BH mass estimates made with the spatial resolution shown
in Figures 1.2 -- 1.3 appear to be reliable.  So how good does the spatial 
resolution have to be?  We can now answer this question
for two $M_\bullet$ analysis machines, the two-integral models of Magorrian et
al.~(1998) and the three-integral models of Gebhardt et al.~(2003).

      Gebhardt et al.~(2003) investigate, for their objects with {\it
      HST\/} spectra and BH detections, how the BH mass would be
affected if only the supporting ground-based observations were 

\begin{figure*}[ht!] % [hb!]
\centering
\vspace{7.5cm}
\includegraphics{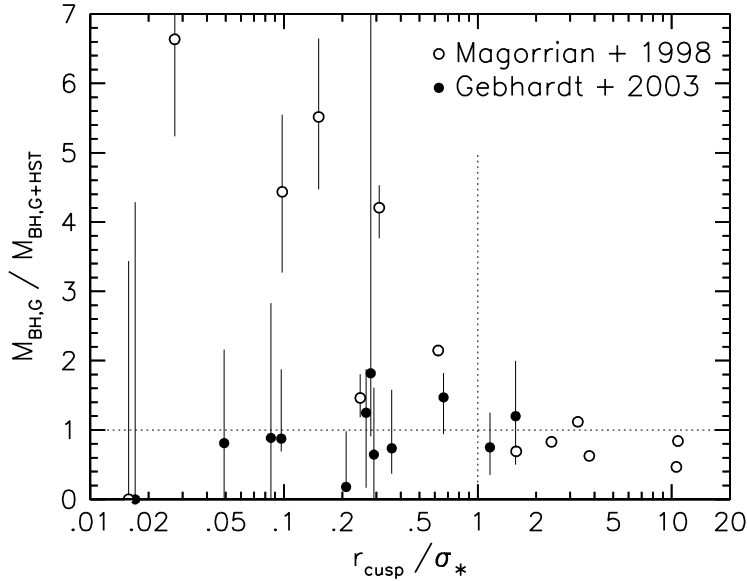}
\vskip 0pt
\caption{Reliability and precision of BH masses as a function of the spatial
resolution of the observations.  The ordinate is the ratio of the BH mass
as obtained from ground-based data to that obtained with {\it HST\/} kinematic
data included.  The error bars are from the ground-based data only, because I
want to illustrate how estimated errors grow as resolution deteriorates.}
\label{fig1.4}
\end{figure*}

\noindent
used in the modeling.  The {\it HST\/} data are higher in resolution than the
ground-based data by a factor of 11.2 $\pm$ 1.2.  If the ground-based
observations are comparable in quality to {\it HST\/} observations made
with the same effective spatial resolution $r_{\rm cusp}/\sigma_*$,
then Gebhardt's exercise distills a clean test of the effects of
spatial resolution.  Modeling uncertainties are minimized because the
same analysis machinery is used on both sets of data.  Gebhardt et
al.~(2003) conclude that, when the {\it HST\/} data are omitted, the error bars
on $M_\bullet$ are larger but the systematic errors in $M_\bullet$ are small.
Here we ask how these results depend on $r_{\rm cusp}/\sigma_*$.

      Figure 1.4 shows no systematic errors in the $M_\bullet$ values given 
by three-integral models, even at low resolution.  BH masses are accurate to a
factor of 1.5 or better provided that $r_{\rm cusp}/\sigma_*$ \gapprox \ts0.3.
All BH detections in Table 1.1 satisfy this criterion.  At lower resolution,
$M_\bullet$ can be wrong by a factor of 2 or more, but the error bars remain
realistic.

      The $M_\bullet$ measurements in Magorrian et al.~(1998) have two main
limitations; they are based on two-integral models, and they are derived from
low-resolution, ground-based spectroscopy.  They can be tested with {\it HST\/}
spectroscopy and (mostly) three-integral models for 13 galaxies (open circles).
When $r_{\rm cusp}/\sigma_*$ \gapprox \ts1, the two-integral models work well;
they underestimate the best current BH masses by a factor of $0.76 \pm 0.09$.
But when $r_{\rm cusp}/\sigma_* < 1$, they overestimate the BH mass by larger
factors at lower resolution.  The reason for the systematic error is unclear.
At $r_{\rm cusp}/\sigma_*$ \lapprox \ts0.1, $M_\bullet$ is overestimated by a
factor of $\sim 5$.  The majority of the Magorrian galaxies that have not been
reobserved with {\it HST\/} are more distant than the ones represented in
Figure~1.4.  Therefore poor resolution plus the assumption of two-integral
models appear to be the reasons why the ratio of BH mass to bulge mass found by
Magorrian et al.~(1998) is larger than the current value of 0.0013 by a factor 
of 4.

      All ground-based BH discoveries in Kormendy \& Richstone (1995)
had $r_{\rm cusp}/\sigma_* > 1$ except in the earliest papers on M{\ts}32.
These papers also overestimated $M_\bullet$ (Fig.~1.1).  {\it HST\/} BH discoveries
made with $r_{\rm cusp}/\sigma_* \simeq 0.3$ to 1 are more secure than the 
M{\ts}32 results derived at the same resolution because we now fit full LOSVDs
and because three-integral models are more reliable than simpler models.

     Given these tests and the ones in Gebhardt (2003), it seems
entirely appropriate that the emphasis in current work has shifted from the
reliability of BH discovery to the use of BH demographics to study the
relationship between BH growth and galaxy formation.

\section{Are They Really Black Holes?}

      Astrophysical arguments that the dark objects detected in galaxy centers
are not clusters of underluminous stars (Maoz 1995, 1998) are well known.
Dark clusters made of brown dwarf stars become luminous when the stars collide, 
merge, and become massive enough for nuclear energy generation.  Clusters of
stellar remnants (white dwarf stars, neutron stars, or stellar-mass black holes)
evaporate as a result of two-body relaxation.  The time scales for these
processes are compellingly short (i.e., \lapprox \ts$10^9$ yr) only for the
Milky Way and for NGC 4258.  The next best case has been M\ts32 (van der Marel
et al.~1998b), although Maoz argued that it is not conclusive.  News in this
subject involves our Galaxy and M{\ts}31.  

      As discussed in \S\ts1.3.2, the observation of an almost-complete, closed
orbit for star S2 in the Sgr A* cluster restricts the central dark mass to live
inside the orbit's pericenter radius, $r_{\rm peri} = 1790$ Schwarzschild
radii.  Demise time scales for dark star clusters are now $< 10^6$ yr.  Even
neutrino balls are excluded (Sch\"odel et al.~2002; Ghez 2003; Ghez et
al.~2003).

      Second, M\ts31 becomes the third galaxy in which astrophysical arguments
make a strong case against dark star clusters.  Bender et al.~(2003) have used
the {\it HST\/} STIS to measure the velocity dispersion of the tiny cluster of blue
stars (King, Stanford, \& Crane 1995; Lauer et al.~1998; Kormendy \& Bender
1999) embedded in the fainter of the two nuclei (``P2'') of the galaxy (Lauer
et al.~1993).
Kormendy \& Bender (1999) already suggested that the central dark object in
M\ts31 is embedded in this blue cluster.~The STIS spectra now show that the 
velocity dispersion of the blue cluster is $\sigma = 940 \pm 100$ km s$^{-1}$
(Fig.~1.5).  This is remarkably high; the red stars along the same line of 
sight have a velocity dispersion of only 300 to 400 km s$^{-1}$.  We can now
be sure that the dark object is in the blue cluster.

      From WFPC2 photometry in Lauer et al.~(1998), the half-light radius of
the blue cluster is $r_h \simeq 0\sd06$.  Since all of the light of the A-type
stars comes from this cluster, $r_h$ and not the {\it HST\/} PSF or 
slit defines the effective spatial resolution of the spectroscopy (Table 1.1).
To confine the stars within the blue cluster, the dark object must have a radius
\hbox{$r_\bullet$ \lapprox \ts$r_h$.}  Also, $M_\bullet$ is larger than we
thought: the virial theorem gives $M_\bullet \approx 2 \times 10^8$ $M_\odot$.
This approximation is an overestimate if the light in the blue cluster is very 
centrally concentrated.  However, it is likely that $M_\bullet$ is at least
7 $\times$ $10^7$ $M_\odot$.  This is the value adopted in Table 1.1.

\begin{figure*}[hb!]
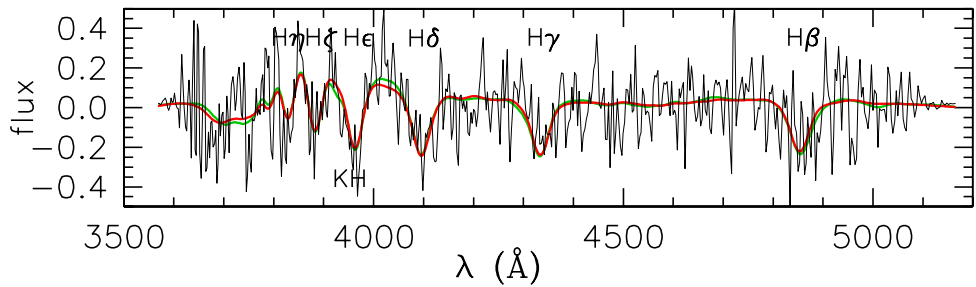
 % [hb!]
\centering
\vspace{4cm}
\bw{ \includegraphics{kormendy_spectrum.ps}}
\col{\includegraphics{kormendy_spectrum.cps}}
\vskip 0pt
\caption{Spectrum (thin line) of the central 0\sd2 of the blue cluster.
The adjacent spectrum of the stars in the bulge and nucleus has been
subtracted.  The spectrum is continuum-subtracted and normalized for the 
Fourier correlation quotient program (Bender 1990).  Flux is in arbitrary 
linear units.  The blue cluster has an A-type spectrum.
Heavy lines show the spectra of an A0 V star and an A0 III star
broadened to the line-of-sight velocity distribution that best fits the
cluster spectrum.  This figure is from Bender et al.~(2003).
}
\label{fig1.5}
\end{figure*}

%
%\pretolerance = 10000 \tolerance = 1000
%\lineskip=0pt  \lineskiplimit=0pt

     It is important to note that $M_\bullet$ is more uncertain in M{\ts}31 than
in other galaxies even though this is the second-nearest BH case.  The
reason is the double~nucleus. Three-integral models are not available and would
omit important physics.  Four techniques have been used. (1) Axisymmetric models
of P1 and P2 give $M_\bullet \simeq ({\rm 4~to~8}) \times 10^7$ $M_\odot$. (2)
Models of the double nucleus as an eccentric disk of stars give $M_\bullet \sim
7 \times 10^7$ $M_\odot$ (Tremaine 1995). (3) The requirement that the center of
mass of the BH and the asymmetric distribution of stars be at the center of the
bulge gives $M_\bullet \sim 3 \times 10^7$ $M_\odot$ (Kormendy \& Bender 1999).
(4) The virial theorem applied to the blue cluster gives $M_\bullet \approx 
2 \times 10^8$ $M_\odot$.  These masses range over a factor of 7.  However, all
four techniques are more uncertain than three-integral models applied to nearly
axisymmetric galaxies.  An improved eccentric disk model has just become available; it gives $M_\bullet \simeq 1 \times 10^8$ $M_\odot$ (Peiris \&
Tremaine 2003).  The most accurate BH mass is likely to come from such detailed
analysis of the asymmetric nucleus.  Here, I adopt a BH mass in the middle of
the above range; it should be accurate to a factor of $\sim 2$. 

      We can now ask: Can we stuff $10^8$ $M_\odot$ of brown dwarfs or stellar
remnants into the central 0\sd06 without getting into trouble?\ts~The answer is
``no'' (Kormendy et al.~2003b).\ts~Following Maoz (1995, 1998), brown
dwarfs are strongly excluded.  The collision time for even the most massive 
brown dwarf (which becomes a luminous star after only one merger) is less than
$10^9$ yr.  Less massive brown dwarfs collide more quickly.  Dark clusters made
of stellar-mass BHs or neutron stars evaporate in several billion years and are
at least weakly excluded.  The most viable dark cluster would be made of
0.6 $M_\odot$ white dwarfs.  Such a cluster would have an evaporation time of
$10^{10}$ yr and is not excluded by the arguments made so far.

      However, we can add a new argument.  An MDO made of stellar remnants
is viable only if its progenitor stars can safely live their lives and deliver
their remnants at suitable radii.  Progenitors get into more trouble than
their remnants.  They are so close together that they collide too quickly,
as follows.  The progenitor cluster must be as small as the dark cluster,
because dynamical friction is too slow to deliver remnants from much larger
radii.  We get into less trouble with collisions if fewer progenitors are
resident at one time.  That is, if the dark cluster was made in time $T$,
and if the progenitor star lifetime is $T_*$, the safest strategy is to have
$T/T_*$ successive generations, each with an equal number of progenitors.  For
$T = 10^{10}$~yr, we then calculate the
time scale on which any one progenitor star collides with another as a function
of the stellar mass and generation number.  The longest time scales are
$10^8$ yr for black hole and neutron star progenitors and shorter for the more
troublesome white dwarf progenitors.  Colliding stars merge and become
progenitors of higher-mass remnants.   Also, stellar mergers decrease the number
of stars and increase the mass range and so shorten the dynamical evolution
time.  The result is a dark cluster with a short evaporation time. 

     So astrophysically plausible alternatives to a supermassive BH are likely
to fail.  Our leverage on the M{\ts}31 BH, $r_{\rm cusp}/\sigma_* \simeq 81$,
is almost as good as $r_{\rm cusp}/\sigma_* \simeq 93$ for NGC 4258.
Astrophysical arguments against BH alternatives are stronger in the latter case
because we know in NGC 4258 but not in M{\ts}31 that the rotation curve is
accurately Keplerian at $r$ \gapprox \ts$\sigma_*$.
This leads to a factor-of-ten stronger constraint on the half-mass radius
of the dark object in NGC 4258 (Maoz 1995, 1998).  Nevertheless, M{\thinspace}31
becomes the third galaxy in which astrophysical arguments favor the conclusion
that a dynamically detected dark object is a BH.  This increases our confidence
that all of them are BHs.  

      Finally, a great variety of AGN observations, including relativistic jets
and X-ray Fe K$\alpha$ line widths as large as 1/3 of the speed of light, argue
forcefully that the engines for nuclear activity in galaxies are BHs.

\section{Conclusion}

    The sanity checks that were the purpose of this paper have succeeded.
Progress on a broad front is on the agenda for this meeting.  It is
gratifying to see the developing connection between the dynamical
BH search and the AGN work that motivated it.  The limited contact
between these subjects was a complaint in Kormendy \& Richstone (1995).  Now, 
reverberation mapping (Blandford \& McKee 1982; Netzer \& Peterson 1997) has
become a reliable tool to measure BH masses (Gebhardt et al.~2000c; Barth 
2003).  Ionization models of AGNs (Netzer 1990; Rokaki, Boisson, \& Collin-Souffrin 1992) are consistent with other techniques (McLure \&
Dunlop 2001; Wandel 2002; Shields et al.~2003).  The growing connection between
BH dynamical searches, AGN physics, and the study of galaxy formation is a sign
of the developing maturity of this subject (e.g., Kormendy 2000).  The
emphasis on BH discovery has given way to the richer field of BH
astrophysics.

      Kormendy \& Richstone (1995) was entitled ``Inward Bound:~The Search for
Supermassive Black Holes in Galaxy Nuclei'' because the BH search is an
iterative process.  ``We make incremental improvements in spatial resolution,
each expensive in ingenuity and money.  [The above] paper reviews the first
order of magnitude of the inward journey in radius.'' At that time, the best
BH candidate, NGC 4258, was observed with a resolution $\sigma_* \simeq 44,000$
Schwarzschild radii (Miyoshi et al.~1995).  Now $\sigma_* \simeq 23,000$
Schwarzschild radii for M{\ts}31 and for NGC 3115.  The best BH case, our
own Galaxy, has $\sigma_* \simeq 1790$ Schwarzschild radii.  The gap between 
the smallest radii reached by dynamical studies and the radii studied by the
well-developed industry on accretion disk physics is shrinking.  This, too, 
is a sign of the growing maturity of the subject.

      But it is too early to declare the problem solved.  Loren Eiseley (1975)
wrote:

{\narrower

``The universe [may be] too frighteningly queer to be understood by
minds like ours.  It's not a popular view.  One is supposed to flourish
Occam's razor and reduce hypotheses about a complex world to human
proportions.  Certainly I try.  Mostly I come out feeling that whatever
else the universe might be, its so-called simplicity is a trick.  I know
that we have learned a lot, but the scope is too vast for us.  Every now and 
then if we look behind us, everything has changed.  It isn't precisely that
nature tricks us.  We trick ourselves with our own ingenuity.''

}

\noindent However reassured we may be by the tests reviewed here,
it is worth remembering that even star S2 in the Galaxy's Sgr A* cluster,
which approaches to within 1790 Schwarzschild radii of the central engine,
lives well outside the region of strong gravity.  Surprises are not out of 
the question.  Further tests of the BH paradigm are worthwhile to make sure
that we do not suddenly find ourselves in an unfamiliar landscape.

\vspace{.5cm}
{\bf Acknowledgements}.
      It is a pleasure to thank my collaborators, R.~Bender, G.~Bower,
the Nuker team
(D.~Richstone, PI), and the STIS GTO team (R. Green, PI) for many helpful
discussions and for permission to discuss our results before publication. 
I am grateful to Luis Ho for his invitation to present this review and for his
patience and meticulous editing.  Careful reading by Scott Tremaine and the
referee resulted in important improvements to this paper.  My {\it HST\/} work
on BHs is supported by grants GO-06587.07, GO-07388.07, \hbox{GO-08591.09-A,}
GO-08687.01-A, and GO-09107.07-A. 

\vfill\eject


\begin{thereferences}{}

\bibitem{} Anders, S.~W., Thatte, N., \& Genzel, R.~2001, in Black Holes in
           Binaries and Galactic Nuclei, ed.~L.~Kaper, E.~P.~J.~van den Heuvel,
           \& P.~A.~Woudt (New York: Springer), 88
%\bibitem{} Andredakis, Y.~C., Peletier, R.~F., \& Balcells, M.~1995, MNRAS,
%     275, 874
%\bibitem{} Andredakis, Y.~C., \& Sanders, R.~H.~1994, MNRAS, 267, 283
%\bibitem{} Axon, D.~J.~2003, in Carnegie Observatories Astrophysics Series,
%           Vol.~1: Coevolution of Black Holes and Galaxies, ed.~L.~C.~Ho
%           (Cambridge: Cambridge Univ. Press), 000
\bibitem{} Bacon, R., Emsellem, E., Combes, F., Copin, Y., Monnet, G., \&
           Martin, P.~2001, A\&A, 371, 409
\bibitem{} Bacon, R., Emsellem, E., Monnet, G., \& Nieto, J.-L.~1994, A\&A, 
           281, 691 % M31
\bibitem{} Backer, D.~C., ed.~1987, The Galactic Center (New York: 
           Amer.~Inst.~Phys.)
%\bibitem{} Balcells, M., \& Peletier, R.~F.~1994, AJ, 107, 135
\bibitem{} Barth, A.~J.~2003, in Carnegie Observatories Astrophysics Series,
           Vol.~1: Coevolution of Black Holes and Galaxies, ed.~L.~C.~Ho
           (Cambridge: Cambridge Univ. Press), 000
\bibitem{} Barth, A.~J., Sarzi, M., Rix, H.-W., Ho, L.~C., Filippenko, A.~V.,
           \& Sargent, W.~L.~W.~2001, ApJ, 555, 685
%\bibitem{} Begelman, M.~C., Blandford, R.~D., and Rees, M.~J.~1980, Nature, 
%           287, 307
\bibitem{} Bender, R.~1990, A\&A, 229, 441 % FCQ
%\bibitem{} Bender, R., Kormendy, J., Bower, G., Green, R., Gull, T., 
%           Hutchings, J.~B., Joseph, C.~L., Kaiser, M.~E., Nelson, C.~H., \&
%           Weistrop, D.~2003, ApJ, submitted 
\bibitem{} Bender, R., et al.~2003, ApJ, submitted 
\bibitem{} Bender, R., Kormendy, J., \& Dehnen, W.~1996, ApJ, 464, L123
\bibitem{} Binney, J.~1978, MNRAS, 183, 501
\bibitem{} Binney, J., \& Mamon, G.~A.~1982, MNRAS, 200, 361
\bibitem{} Blandford, R.~D., \& McKee, C.~F.~1982, ApJ, 255, 419
%\bibitem{} Blandford, R.~D., McKee, C.~F., \& Rees, M.~J.~1977, Nature, 
%      267, 211
%\bibitem{} B\"oker, T., van der Marel, R.~P., \& Vacca, W.~D.~1999, AJ, 118, 831
%\bibitem{} Bower, G.~A., Green, R.~F., Bender, R., Gebhardt, K., Lauer, T.~R.,
%           Magorrian, J., Richstone, D.~O., Danks, A., Gull, T., et al.~2001,
%           ApJ, 550, 75
\bibitem{} Bower, G.~A., et al.~1998, ApJ, 492, L111  %  NGC 4374
\bibitem{} ------.~2001, ApJ, 550, 75
%\bibitem{} Bower, G.~A., Green, R.~F., Danks, A., Gull, T., Heap, S., 
%           Hutchings, J., Joseph, C., Kaiser, M.~E., Kimble, R., et al.~ 1998,
%           ApJ, 492, L111  %  NGC 4374
\bibitem{} Bower, G.~A., Wilson, A.~S., Heckman, T.~M., Magorrian, J., 
           Gebhardt, K., Richstone, D.~O., Peterson, B.~M., \& 
           Green, R.~F.~2000, BAAS, 32, 1566
%\bibitem{} Bridle, A.~H., Perley, R.~A.~1984, ARA\&A, 22, 319
%\bibitem{} Brown, T.~M., et al.~1998, AJ, 504, 113
\bibitem{} Cappellari, M., Verolme, E.~K., van der Marel, R.~P., 
           Verdoes Kleijn, G.~A., Illingworth, G.~D., Franx, M., Carollo, C.~M.,
           \& de Zeeuw, P.~T.~2002, ApJ, 578, 787
%\bibitem{} Carollo, C.~M., \& Stiavelli, M.~1998a, AJ, 115, 2306
%\bibitem{} Carollo, C.~M., Stiavelli, M., de Zeeuw, P.~T., \& Mack, J.~1997, 
%           AJ, 114, 2366
%\bibitem{} Carollo, C.~M., Stiavelli, M., \& Mack, J.~1998b, AJ, 116, 68
\bibitem{} Chakrabarty, D., \& Saha, P.~2001, AJ, 122, 232
\bibitem{} Chary, R., et al.~2000, ApJ, 531, 756
%\bibitem{} Combes, F.~2000, in Galaxy Disks and Disk Galaxies, ed.~J.~G.~Funes
%          \& E.~M.~Corsini (San Francisco: PASP), in press      
%\bibitem{} Combes, F.~2000, these proceedings     
%\bibitem{} Combes, F., Debbasch, F., Friedli, D. \& Pfenniger, D.~1990, A\&A,
%           233, 82
%\bibitem{} Courteau, S., de Jong, R.~S., \& Broeils, A.~H.~1996, ApJ, 457, L73
\bibitem{} Cretton, N., Copin, Y., Emsellem, E., \& de Zeeuw, T.~2003, in 
Carnegie Observatories Astrophysics Series, Vol. 1: Coevolution of Black Holes 
and Galaxies, ed. L. C. Ho (Pasadena: Carnegie Observatories,
http://www.ociw.edu/ociw/symposia/series/symposium1/proceedings.html)
\bibitem{} Cretton, N., de~Zeeuw, P.~T., van~der~Marel, R.~P., \& Rix, 
           H.-W.~1999b, ApJS, 124, 383   
\bibitem{} Cretton, N., \& van den Bosch, F.~C.~1999a, ApJ, 514, 704
\bibitem{} Dehnen, W.~1995, MNRAS, 274, 919 
\bibitem{} Devereux, N., Ford, H., Tsvetanov, Z., \& Jacoby, G.~2003,
           AJ, 125, 1226 
\bibitem{} Dressler, A., \& Richstone, D.~O.~1988, ApJ, 324, 701
\bibitem{} ------.~1990, ApJ, 348, 120
\bibitem{} Duncan, M.~J., \& Wheeler, J.~C.~1980, ApJ, 237, L27
%\bibitem{} Ebisuzaki, T., Makino, J., Okamura, S.~K.~1991, Nature, 354, 212
\bibitem{} Eckart, A., \& Genzel, R.~1997, MNRAS, 284, 576
\bibitem{} Eiseley, L.~1975, All the Strange Hours (New York: Scribner)
\bibitem{} Emsellem, E., Dejonghe, H., \& Bacon, R.~1999, MNRAS, 303, 495
\bibitem{} Emsellem, E., Monnet, G., Bacon, R., \& Nieto, J.-L.~1994, 
           A\&A, 285, 739
\bibitem{} Evans, N.~W., \& de Zeeuw, P.~T.~1994, MNRAS, 271, 202 
%\bibitem{} Faber, S.~M., \& Jackson, R.~E.~1976, ApJ, 204, 668
%\bibitem{} Faber, S.~M., Tremaine, S., Ajhar, E.~A., Byun, Y.-I., Dressler, A.,
%           Gebhardt, K., Grillmair, C., Kormendy, J., Lauer, T.~R., \&
%           Richstone, D.~1997, AJ, 114, 1771
\bibitem{} Faber, S.~M., et al.~1997, AJ, 114, 1771 % Nuker IV
%\bibitem{} Ferrarese, L.~2002, astro-ph/0203047  % Zoltan argument
\bibitem{} Ferrarese, L., \& Ford, H.~C.~1999, ApJ, 515, 583
\bibitem{} Ferrarese, L., Ford, H.~C., \& Jaffe, W.~1996, ApJ, 470, 444
%\bibitem{} Ferrarese, L., Merritt, D.~2000,  ApJ, 539, L9 
%\bibitem{} Filippenko, A.~V.~(ed.)~1992, Relationships Between Active Galactic
%      Nuclei and Starburst Galaxies (San Francisco: ASP)
%\bibitem{} Filippenko, A.~V., Ho, L.~2003, preprint
%\bibitem{} Freeman, K.~C.~1970, ApJ, 160, 811
%
\bibitem{} Gebhardt, K.~2003, in Carnegie Observatories Astrophysics Series,
           Vol.~1: Coevolution of Black Holes and Galaxies, ed.~L.~C.~Ho
           (Cambridge: Cambridge Univ. Press), 000
\bibitem{} Gebhardt, K. et al. 2000a, AJ, 119, 1157 % NGC 3379
\bibitem{} ------. 2000b, ApJ, 539, L13 % BH-sigma correlation
\bibitem{} ------. 2000c, ApJ, 543, L5  % Reverberation BH masses
%\bibitem{} Gebhardt, K., et al.~2001, AJ, 122, 2469 % M33
\bibitem{} ------.~2003, ApJ, 583, 92
\bibitem{} 
Gebhardt, K., Richstone, D., Ajhar, E.~A., Kormendy, J., Dressler, A., Faber,
S.~M., Grillmair, C., \& Tremaine, S. 1996, AJ, 112, 105 % Nuker III
%
%\bibitem{} Gebhardt, K., Bender, R., Bower, G., Dressler, A., Faber, S.~M.,
%           Filippenko, A.~V., Green, R., Grillmair, C., Ho, L.~C., 
%           et al.~2002a, ApJ, 539, L13 
%\bibitem{} Gebhardt, K., Lauer, T.~R., Kormendy, J., Pinkney, J., Bower, G.~A.,
%           Green, R., Gull, T., Hutchings, J.~B., Kaiser, M.~E., et al.~2001,
%           AJ, 122, 2469 % M33
%\bibitem{} Gebhardt, K., Kormendy, J., Ho, L.~C., Bender, R., Bower, G.,
%           Dressler, A., Faber, S.~M., Filippenko, A.~V., Green, R., 
%           et al.~2002b, ApJ, 543, L5 
%\bibitem{} Gebhardt, K., Richstone, D., Kormendy, J., Lauer, T.~R., 
%           Ajhar, E.~A., Bender, R., Dressler, A., Faber, S.~M., Grillmair, C.,
%           et al.~2000a, AJ, 119, 1157 % NGC 3379
%\bibitem{} Gebhardt, K., Richstone, D., Tremaine, S., Lauer, T.~R., 
%           Bender, R., Bower, G., Dressler, A., Faber, S.~M., 
%           Filippenko, A.~V., et al.~2003, ApJ, 583, 92
%
\bibitem{} Genzel, R., Eckart, A., Ott, T., \& Eisenhauer, F.~1997, MNRAS, 
           291, 219
\bibitem{} Genzel, R., \& Harris, A.~I., ed.~1994, The Nuclei of Normal
           Galaxies: Lessons From The Galactic Center (Dordrecht: Kluwer)
\bibitem{} Genzel, R., Hollenbach, D., \& Townes, C.~H.~1994, Rep.~Prog. ~Phys.,
           57, 417
%\bibitem{} Genzel, R., Lutz, D., Sturm, E., Egami, E., Kunze, D., 
%           Moorwood, A.~F.~M., Rigopoulou, D., Spoon, H.~W.~W., 
%           Sternberg, A., et al.~1998, ApJ, 498, 579
%\bibitem{} Genzel, R., et al.~1998, ApJ, 498, 579
\bibitem{} Genzel, R., Pichon, C., Eckart, A., Gerhard, O.~E., \& Ott, T.~2000,
           MNRAS, 317, 348
\bibitem{} Genzel, R., Thatte, N., Krabbe, A., Kroker, H., \& 
           Tacconi-Garman, L.~E.~1996, ApJ, 472, 153
\bibitem{} Genzel, R., \& Townes, C.~H.~1987, ARA\&A, 25, 377
%\bibitem{} Gerhard, O.~E., \& Binney, J.~1985, MNRAS, 216, 467
\bibitem{} Ghez, A.~M.~2003, in Carnegie Observatories Astrophysics Series,
           Vol.~1: Coevolution of Black Holes and Galaxies, ed.~L.~C.~Ho
           (Cambridge: Cambridge Univ. Press), 000
\bibitem{} Ghez, A.~M., et al.~2003, ApJ, in press (astro-ph/0302299)
\bibitem{} Ghez, A.~M., Klein, B.~L., Morris, M., \& Becklin, E.~E.~1998, 
           ApJ, 509, 678
%\bibitem{} Ghez, A.~M., et al.~2000, in Black Holes in Binaries and 
%           Galactic Nuclei (Garching: ESO), in press
%\bibitem{} Ghez, A.~M., et al.~2000, Nature, 407, 349
%\bibitem{} Green, R.~F., et al.~2001, in preparation
\bibitem{} Greenhill, L.~J., \& Gwinn, C.~R.~1997a, Ap\&SS, 248, 261 % N1068
%\bibitem{} Greenhill, L.~J., Gwinn, C.~R., Antonucci, R., Barvainis, R.,
%           1996, ApJ, 472, L21 % N1068 not adopted (use ApSS article)
\bibitem{} Greenhill, L.~J, Moran, J.~M., \& Herrnstein, J.~R.~1997b, 
           ApJ, 481, L23 % N4945 BH mass
\bibitem{} Haller, J.~W., Rieke, M.~J., Rieke, G.~H., Tamblyn, P., Close, L., 
           \& Melia, F.~1996, ApJ, 456, 194
%\bibitem{} Harms, R.~J., Ford, H.~C., Tsvetanov, Z.~I., Hartig, G.~F., 
%           Dressel, L.~L., Kriss, G.~A., Bohlin, R., Davidsen, A.~F., 
%           et al.~ 1994, ApJ, 435, L35
\bibitem{} Harms, R.~J., et al.~ 1994, ApJ, 435, L35
%\bibitem{} Hasan, H., Pfenniger, D., \& Norman, C.~1993, ApJ, 409, 91
\bibitem{} Herrnstein, J.~R., Moran, J.~M., Greenhill, L.~J., Diamond, P.~J.,
           Inoue, M., Nakai, N., Miyoshi, M., Henkel, C., \& Riess, A.~1999,
           Nature, 400, 539
\bibitem{} Ho, L.~C.~1999, in Observational Evidence for Black Holes in 
           the Universe, ed.~S.~K.~Chakrabarti (Dordrecht: Kluwer), 157
%\bibitem{} Ho, L.~C.~2003, Carnegie Observatories Astrophysics Series, Vol.~1:
%           Coevolution of Black Holes and Galaxies, ed.~L.~C.~Ho (Cambridge: 
%           Cambridge Univ. Press
%\bibitem{} Ho, L.~C., Rudnick, G., Rix, H.-W., Shields, J.~C., McIntosh, D.~H.,
%           Filippenko, A.~V., Sargent, W.~L.~W., \& Eracleous, M.~2000, \apj,
%           541, 120
%\bibitem{} Holley-Bockelmann, K., Mihos, C., Sigurdsson, S., Hernquist, L., 
%           \& Norman, C.~A.~2001, ApJ, submitted
%\bibitem{} Holley-Bockelmann, K., Richstone, D.~1999, ApJ, 517, 92
%\bibitem{} Holley-Bockelmann, K., Richstone, D.~2000, ApJ, 531, 232
\bibitem{} Illingworth, G.~1977, ApJ, 218, L43
%\bibitem{} Ivison, R.~J., et al.~2000, MNRAS, 315, 209
%\bibitem{} Jones, D.~H., et al.~ 1996, ApJ, 466, 742
%\bibitem{} Joseph, C.~L., Merritt, D., Olling, R., Valluri, M., Bender, R.,
%           Bower, G., Danks, A., Gull, T., Hutchings, J.~B., et al.~2001, 
%           ApJ, 550, 668
\bibitem{} Joseph, C.~L., et al.~2001, ApJ, 550, 668
%\bibitem{} Joseph, R.~D.~1999, A\&SS, 266, 321
\bibitem{} Kaiser, M.~E., et al.~2003, in preparation
\bibitem{} Kent, S.~M.~1992, ApJ, 387, 181
%\bibitem{} King, I.~R., Deharveng, J.~M., Albrecht, R., Barbieri, C., 
%           Blades, J.~C., Boksenberg, A., Crane, P., Disney, M.~J., 
%           Jakobsen, P., et al.~1992, ApJ, 397, L35
\bibitem{} King, I.~R., Stanford, S.~A., \& Crane, P.~1995, AJ, 109, 164
%\bibitem{} Kormendy, J.~1985, ApJ, 295, 73
%\bibitem{} Kormendy, J.~1987, in IAU Symposium 127, Structure and Dynamics of
%           Elliptical Galaxies, ed.~T.~de Zeeuw (Dordrecht: Reidel), 17
%\bibitem{} Kormendy, J.~1987, in Nearly Normal Galaxies: From the Planck Time 
%           to the Present, ed.~S.~M.~Faber (New York: Springer-Verlag), 163
%\bibitem{} Kormendy, J.~1988a, in Supermassive Black Holes, ed. M. Kafatos
%           (Cambridge: Cambridge Univ.~Press), 98
\bibitem{} Kormendy, J.~1988a, ApJ, 325, 128
\bibitem{} ------.~1988b, ApJ, 335, 40
%\bibitem{} Kormendy, J.~1989, ApJ, 342, L63
%\bibitem{} Kormendy, J.~1993a, in The Nearest Active Galaxies, ed.~J.~Beckman,
%     L.~Colina, \& H.~Netzer (Madrid: Consejo Superior de Investigaciones
%     Cient\'\i ficas), 197
%\bibitem{} Kormendy, J.~1993a, in IAU Symposium 153, Galactic Bulges,
%           eds.~H.~Dejonghe \& H.~Habing (Dordrecht: Kluwer), 209
\bibitem{} ------.~1993, in The Nearest Active Galaxies, eds.~J.~Beckman,
           L.~Colina, \& H. Netzer (Madrid: Consejo Superior de Investigaciones
           Cient\'\i ficas), 197 
\bibitem{} ------.~2000, Science, 289, 1484
\bibitem{} Kormendy, J., \& Bender, R.~1996, ApJ, 464, L119
\bibitem{} ------.~1999, ApJ, 522, 772 % (KB)
\bibitem{} Kormendy, J., \& Bender, R., Evans, A.~S., \& Richstone, D.~1998, 
           AJ, 115, 1823
\bibitem{} Kormendy, J., Bender, R., Freeman, K.~C., Ambrose, E., \&
           Tonry, J.~L.~2003a, in preparation 
\bibitem{} Kormendy, J., Dressler, A., Byun, Y.-I., Faber, S.~M., 
           Grillmair, C., Lauer, T.~R., Richstone, D., \& Tremaine, S.~1994, 
           in ESO/OHP Workshop on Dwarf Galaxies, ed.~G.~Meylan \& P.~Prugniel
           (Garching: ESO), 147
\bibitem{} Kormendy, J., \& Gebhardt, K.~2001, in 20$^{\rm th}$ Texas
      Symposium on Relativistic Astrophysics, ed.~J.~C.~Wheeler \& H.~Martel
      (Melville: AIP), 363
\bibitem{} Kormendy, J., \& Richstone, D.~1992, ApJ, 393, 559
\bibitem{} ------.~1995, ARA\&A, 33, 581 
%\bibitem{} Kormendy, J., \& Sanders, D.~B.~1992, ApJ, 390, L53
%\bibitem{} Kormendy, J., \& Westpfahl, D.~J.~1989, ApJ, 338, 752
%
%
\bibitem{} Kormendy, J., et al.~1996a, ApJ, 459, L57  %  NGC 3115
\bibitem{} ------.~1996b, ApJ, 473, L91  %  NGC 4594
\bibitem{} ------.~1997, ApJ, 482, L139  %  NGC 4486B
\bibitem{} ------.~2003b, in preparation
%
%
%\bibitem{} Kormendy, J., Bender, R., Ajhar, E.~A., Dressler, A., Faber, S.~M.,
%           Gebhardt, K., Grillmair, C., Lauer, T.~R., Richstone, D., 
%           et al.~1996b, ApJ, 473, L91 
%\bibitem{} Kormendy, J., Bender, R., Magorrian, J., Tremaine, S., Gebhardt, K.,
%           Richstone, D., Dressler, A., Faber, S.~M., Grillmair, C., 
%           et al.~1997, ApJ, 482, L139  %  NGC 4486B
%\bibitem{} Kormendy, J., Bender, R., Richstone, D., Ajhar, E.~A.,
%           Dressler, A., Faber, S.~M., Gebhardt, K., Grillmair, C., 
%           Lauer, T.~R., et al.~1996a, ApJ, 459, L57 
%\bibitem{} Krabbe, A., Genzel, R., Eckart, A., Najarro, F., Lutz, D., 
%           Cameron, M., Kroker, H., Tacconi- Garman, L.~E., Thatte, N., 
%           et al.~1995, ApJ, 447, L95
\bibitem{} Krabbe, A., et al.~1995, ApJ, 447, L95
%\bibitem{} Lake, G., \& Norman, C.~1983, ApJ, 270, 51
%\bibitem{} Laor, A.~1998, ApJ, 505, L83  
\bibitem{} Lauer, T.~R.~2003, in Carnegie Observatories Astrophysics Series,
           Vol.~1: Coevolution of Black Holes and Galaxies, ed.~L.~C.~Ho
           (Cambridge: Cambridge Univ. Press), 000
\bibitem{} Lauer, T.~R., et al.~1993, AJ, 106, 1436
\bibitem{} ------.~1995, AJ, 110, 2622 % Nuker I
\bibitem{} 
Lauer, T.~R., Faber, S.~M., Ajhar, E.~A., Grillmair, C.~J., \& Scowen,
P.~A. 1998, \aj, 116, 2263
%\bibitem{} Lauer, T.~R., Gebhardt, K., Richstone, D., Tremaine, S., Bender, R.,
%           Bower, G., Dressler, A., Faber, S.~M., Filippenko, A.~V., 
%           et al.~2002, AJ, 124, 1975
%\bibitem{} Lauer, T.~R., et al.~2002, AJ, 124, 1975
%\bibitem{} Light, E.~S., Danielson, R.~E., \& Schwarzschild, M.~1974, ApJ, 
%           194, 257
%\bibitem{} Lutz, D., et al.~1998, ApJ, 505, L103
\bibitem{} Lynden-Bell, D.~1969, Nature, 223, 690
\bibitem{} ------.~1978, Physica Scripta, 17, 185
\bibitem{} Macchetto, F., Marconi, A., Axon, D.~J., Capetti, A., Sparks, W.,
           \& Crane, P.~1997, ApJ, 489, 579
\bibitem{} Maciejewski, W., \& Binney, J. 2001, MNRAS, 323, 831
%\bibitem{} Magorrian, J., Tremaine, S., Richstone, D., Bender, R., Bower, G.,
%           Dressler, A., Faber, S.~M., Gebhardt, K., Green, R., et al.~1998,
%           AJ, 115, 2285  
\bibitem{} Magorrian, J., et al.~1998, AJ, 115, 2285  
%\bibitem{} Makino, J., Ebisuzaki, T.~1996, ApJ, 465, 527
\bibitem{} Maoz, E.~1995, ApJ, 447, L91
\bibitem{} ------.~1998, ApJ, 494, L181 
\bibitem{} Marconi, A., Capetti, A., Axon, D.~J., Koekemoer, A., Macchetto, D.,
           \& Schreier, E.~J.~2001, ApJ, 549, 915
%\bibitem{} Marconi, A., et al.~ 2001, ApJ, 549, 915
\bibitem{} McLure, R.~J., \& Dunlop, J.~S.~2001, MNRAS, 327, 199
%\bibitem{} Merritt, D.~1999, PASP, 111, 129
%\bibitem{} Merritt, D., \& Quinlan, G.~D.~1998, ApJ, 498, 625
\bibitem{} Merritt, D., \& Ferrarese, L.~2001, MNRAS, 320, L30 %BH/bulge=.0013
%\bibitem{} Merritt, D., Ferrarese, L., \& Joseph, C.~L.~2001, Science, 293,
%           1116 
%\bibitem{} Mihos, J.~C., Hernquist, L.~1994, ApJ, 437, L47
%\bibitem{} Milosavljevi\'c, M., \& Merritt, D.~2001,  ApJ 563, 34 
\bibitem{} Miyoshi, M., Moran, J., Herrnstein, J., Greenhill, L., Nakai, N.,
           Diamond, P., \& Inoue, M.~1995, Nature, 373, 127
\bibitem{} Morris, M., ed.~1989, IAU Symp. 136, The Center of Our Galaxy
           (Dordrecht: Kluwer)
\bibitem{} Munyaneza F., Tsiklauri D., \& Viollier R.~D.~1998, ApJ, 509, L105
\bibitem{} ------.~1999, ApJ, 526, 744
\bibitem{} Munyaneza F., \& Viollier R.~D.~2002, ApJ, 564, 274
%\bibitem{} Nakano, T., \& Makino, J.~1999a, ApJ, 510, 155
%\bibitem{} Nakano, T., \& Makino, J.~1999b, ApJ, 525, L77
%\bibitem{} Nelson, C.~H., et al.~ 2001, in preparation
\bibitem{} Netzer, H.~1990, in Active Galactic Nuclei, Saas-Fee Advanced Course
           20, ed.~T.~J.-L.~Courvoisier \& M.~Mayor (Berlin: Springer), 57 
\bibitem{} Netzer, H., \& Peterson, B.~M.~1997, in Astronomical Time Series,
           ed.~D.~Maoz, A.~Sternberg, \& E.~M.~Leibowitz (Dordrecht: Kluwer), 85
%\bibitem{} Norman, C.~A., May, A., \& van Albada, T.~S.~1985, ApJ, 296, 20
%\bibitem{} Norman, C.~A., Sellwood, J.~A., \& Hasan, H.~1996, ApJ, 462, 114
\bibitem{} Peiris, H.~V., \& Tremaine, S.~2003, ApJ, submitted
%\bibitem{} Peletier, R.~F., et al.~ 2000, MNRAS, 310, 703
%\bibitem{} Perley, R.~A., Dreher, J.~W., \& Cowan, J.~J.~1984, ApJ, 285, L35
%\bibitem{} Peterson, B.~2003, in Carnegie Observatories Astrophysics Series,
%           Vol.~1: Coevolution of Black Holes and Galaxies, ed.~L.~C.~Ho
%           (Cambridge: Cambridge Univ. Press), 000
%\bibitem{} Pfenniger, D., \& Norman, C.~1990, ApJ, 363, 391
%\bibitem{}Pfenniger, D., \& Norman, C.~A.~1991, in IAU Symposium 146, Dynamics 
%          of Galaxies and Their Molecular Cloud Distributions, ed.~F.~Combes \&
%           F.~Casoli (Dordrecht: Kluwer), 323
%\bibitem{} Pickles, A.~J.~1998, PASP, 110, 863
%\bibitem{} Pinkney, J., et al.~ 2003, ApJ, in press
%\bibitem{} Poon, M.~Y., \& Merritt, D.~2001, ApJ, 549, 192
\bibitem{} Qian, E.~E., de Zeeuw, P.~T., van der Marel, R.~P., \& 
           Hunter, C.~1995, MNRAS, 274, 602
%\bibitem{} Quinlan, G.~D.~1996, NewA, 1, 35
%\bibitem{} Quinlan, G.~D., Hernquist, L.~1997, NewA, 2, 533 
%\bibitem{} Quinlan, G.~D., Hernquist, L., \& Sigurdsson, S.~1995, ApJ, 440, 554
%\bibitem{} Rees, M.~J.~1984, ARA\&A, 22, 471
%\bibitem{} Regan, M.~W., et al.~2001, ApJ, 561, 218
%\bibitem{} Reynolds, C.~S., Nowak, M.~A.~2002, Physics Reports, in press,
%           astro-ph/0212065 
\bibitem{} Richstone, D.~2003, in Carnegie Observatories Astrophysics Series,
           Vol.~1: Coevolution of Black Holes and Galaxies, ed.~L.~C.~Ho
           (Cambridge: Cambridge Univ. Press), 000
%\bibitem{} Richstone, D., Ajhar, E.~A., Bender, R., Bower, G., Dressler, A.,
%           Faber, S.~M., Filippenko, A.~V., Gebhardt, K., Green, R.,
%           et al.~1998, Nature, 395A, 14 
\bibitem{} Richstone, D., et al.~1998, Nature, 395, A14 
\bibitem{} ------.~ 2003, in preparation
\bibitem{} Richstone, D., Bower, G., \& Dressler, A.~1990, ApJ, 353, 118
\bibitem{} Richstone, D.~O., \& Tremaine, S.~1985, ApJ, 296, 370 % M87 sans BH
\bibitem{} Rokaki, E., Boisson, C., \& Collin-Souffrin, S.~1992, A\&A, 253, 57
\bibitem{} Salpeter, E.~E.~1964, ApJ, 140, 796
%\bibitem{} Sanders, D.~B.~1999, A\&SS, 266, 331
%\bibitem{} Sanders, D.~B., \& Mirabel, I.~F.~1996, ARA\&A, 34, 749
%\bibitem{} Sanders, D.~B., et al.~1988a, ApJ, 325, 74
%\bibitem{} Sanders, D.~B., Soifer, B.~T., Elias, J.~H., Neugebauer, G., \&
%     Matthews, K.~1988b, ApJ, 328, L35
%\bibitem{} Sanders, D.~B., et al.~1988b, ApJ, 328, L35
\bibitem{} Sargent, W.~L.~W., Young, P.~J., Boksenberg, A., Shortridge, K., 
     Lynds, C.~R., \& Hartwick, F.~D.~A.~1978, ApJ, 221, 731
\bibitem{} Sarzi, M.~2003, in Carnegie Observatories Astrophysics
Series, Vol. 1: Coevolution of Black Holes and Galaxies, ed. L. C. Ho
(Pasadena: Carnegie Observatories,
http://www.ociw.edu/ociw/symposia/series/symposium1/proceedings.html)
\bibitem{} Sarzi, M., Rix, H.-W., Shields, J.~C., Rudnick, G., Ho, L.~C.,
           McIntosh, D.~H., Filippenko, A.~V., \& Sargent, W.~L.~W.~2001, 
           ApJ, 550, 65
%\bibitem{} Schmidt, M.~1963, Nature, 197, 1040
%\bibitem{} Sch\"odel, R., Ott, T., Genzel, R., Hofmann, R., Lehnert, M., 
%           Eckart, A., Mouawad, N., Alexander, T., Reid, M.~J., et al.~2002, 
%           Nature, 419, 694 
\bibitem{} Sch\"odel, R., et al.~2002, Nature, 419, 694
\bibitem{} Schwarzschild, M.~1979, ApJ, 232, 236 
\bibitem{} Sellgren, K., McGinn, M.~T., Becklin, E.~E., \& Hall, D.~N.~B.~1990,
           ApJ, 359, 112
%\bibitem{} Sellwood, J.~A., \& Moore, E.~M.~1999, ApJ, 510, 125
\bibitem{} Shields, G.~A., Gebhardt, K., Salviander, S., Wills, B.~J., Xie, B.,
           Brotherton, M.~S., Yuan, J., \& Dietrich, M.~2003, ApJ, 583, 124
%\bibitem{} Silk, J., \& Rees, M.~J.~1998, A\&A, 331, L1
\bibitem{} Siopis, C., et al.~2003, in preparation
%\bibitem{} Statler, T.~S., King, I.~R., Crane, P., \& Jedrzejewski, R.~I.~1999,
%           AJ, 117, 894
\bibitem{} Tonry, J.~L.~1984, ApJ, 283, L27
\bibitem{} ------.~1987, ApJ, 322, 632
\bibitem{} Tonry, J.~L., et al.~ 2001, ApJ, 546, 681
% \bibitem{} Toomre, A.~1964, ApJ, 139, 1217
\bibitem{} Tremaine, S.~1995, AJ, 110, 628
%\bibitem{} Tremaine, S., Gebhardt, K., Bender, R., Bower, G., Dressler, A.,
%           Faber, S.~M., Filippenko, A.~V., Green, R., Grillmair, C., 
%           et al.~2002, ApJ, 574, 740 
\bibitem{} Tremaine, S., et al.~2002, ApJ, 574, 740 
\bibitem{} Tsiklauri D., \& Viollier R.~D.~1998, ApJ, 500, 591
\bibitem{} ------.~1999, Astroparticle Phys., 12, 199
%\bibitem{} Valluri, M., \& Merritt, D.~1998, ApJ, 506, 686
\bibitem{} Valluri, M., Merritt, D., \& Emsellem, E.~2003, \apj, submitted (astro-ph/0210379)
\bibitem{} van der Marel, R.~P.~1994, ApJ, 432, L91
\bibitem{} van~der~Marel, R.~P., Cretton, N., de Zeeuw, P.~T., \& Rix,
           H.-W.~1998b, ApJ, 493, 613 % M32
\bibitem{} van der Marel, R.~P., de Zeeuw, P.~T., Rix, H.-W.~1997b, ApJ, 
           488, 119
\bibitem{} van der Marel, R.~P., de Zeeuw, P.~T., Rix, H.-W., \&
           Quinlan, G.~D.~1997a, Nature, 385, 610
\bibitem{} van der Marel, R.~P., Evans, N.~W., Rix, H.-W., White, S.~D.~M., 
           \& de Zeeuw, T.~1994b, MNRAS, 271, 99 % M32 models and BH mass
\bibitem{} van der Marel, R.~P., Rix, H.-W., Carter, D., Franx, M., 
           White, S.~D.~M., \& de Zeeuw, P.~T.~1994a, MNRAS, 268, 521
\bibitem{} van der Marel, R.~P., \& van den Bosch, F.~C.~1998a, AJ, 116, 2220
           % N7052
%\bibitem{} Veilleux, S.~2000, astro-ph/0012121
%\bibitem{} Verdoes Kleijn, G.~A., van der Marel, R.~P., Carollo, C.~M., \& 
%           de Zeeuw, P.~T.~2000, AJ, 120, 1221
\bibitem{} Verdoes Kleijn, G.~A., van der Marel, R.~P., de Zeeuw, P.~T.,
           Noel-Storr, J., \& Baum, S.~A.~2002, AJ, 124, 2524
%\bibitem{} Verdoes Kleijn, G.~A., et al.~ 2001, AJ, 120, 1221
\bibitem{} Verolme, E.~K., et al.~2002, MNRAS, 335, 517
%\bibitem{} Wandel, A.~1999, ApJ, 519, L39
\bibitem{} Wandel, A.~2002, ApJ, 565, 762
%\bibitem{} Wandel, A., Peterson, B.~M., \& Malkan, M.~A.~1999, ApJ, 526, 579
\bibitem{} Young, P.~J., Westphal, J.~A., Kristian, J., Wilson, C.~P., \& 
           Landauer, F.~P.~1978, ApJ, 221, 721
%\bibitem{} Yu, Q., \& Tremaine, S.~2002, MNRAS, 335, 965
%\bibitem{} Yusef-Zadeh, F., Melia, F., \& Wardle, M.~2000, Science, 287, 85
\bibitem{} Zel'dovich, Ya.~B.~1964, Soviet Physics -- Doklady, 9, 195



\end{thereferences}
\end{document}